\documentclass{emulateapj}
\usepackage{apjfonts}
\usepackage{lscape}
\newcommand{\rotator}{}
\newcommand{\lscapeclose}{\end{landscape}}
\newcommand{\lscapeopen}{\begin{landscape}}
\newcommand{\sizer}{}
\newcommand{\calculatorurl}{\scriptsize{\path{http://www.cfa.harvard.edu/~phopkins/Site/qlf.html}}}

\newcommand{\Lstar}{L_{\ast}}
\newcommand{\lstar}{\Lstar}
\newcommand{\Pstar}{\phi_{\ast}}

\newcommand{\phistar}{\Pstar}
\newcommand{\slopefaint}{\gamma_{1}}
\newcommand{\slopebright}{\gamma_{2}}

\newcommand{\etal}{et al.}
\newcommand{\reducechi}{\chi^{2}/\nu}
\newcommand{\NH}{N_{\rm H}}
\newcommand{\nh}{\NH}
\newcommand{\lessim}{\lesssim}

\shorttitle{Bolometric Quasar Luminosity Function}
\shortauthors{Hopkins \etal}
\slugcomment{Submitted to ApJ, May 23, 2006}
\begin{document}

\title{An Observational Determination of the 
Bolometric Quasar Luminosity Function}
\author{Philip F. Hopkins\altaffilmark{1}, 
Gordon T. Richards\altaffilmark{2}, 
\&\ Lars Hernquist\altaffilmark{1} 
}
\altaffiltext{1}{Harvard-Smithsonian Center for Astrophysics, 
60 Garden Street, Cambridge, MA 02138}
\altaffiltext{2}{Department of Physics and Astronomy, The Johns 
Hopkins University, 3400 North Charles Street, Baltimore, MD 21218}
\altaffiltext{3}{\calculatorurl}

\begin{abstract}

We combine a large set of quasar luminosity function (QLF)
measurements from the rest-frame optical, soft and hard X-ray, and
near- and mid-infrared bands to determine the bolometric QLF in the redshift
interval $z=0-6$.  Accounting for the observed distributions of quasar
column densities and variation of spectral energy distribution (SED)
shapes, and their dependence on luminosity, makes it possible to
integrate the observations in a reliable manner and provides a
baseline in redshift and luminosity larger than that of any individual
survey.  We infer the QLF break luminosity and faint-end
slope out to $z\sim4.5$ and confirm at
high significance ($\gtrsim10\sigma$)
previous claims of a
flattening in both the faint- and bright-end slopes with redshift.
With the best-fit estimates of
the column density distribution and quasar SED, which both depend on
luminosity, a single bolometric QLF self-consistently reproduces the
observed QLFs in all bands and at all redshifts for which we compile
measurements.  Ignoring this luminosity dependence does not yield a
self-consistent bolometric QLF and there is no evidence for any
additional dependence on redshift.  We calculate the expected relic
black hole mass function and mass density, cosmic X-ray background,
and ionization rate as a function of redshift and find they are
consistent with existing measurements. The peak in the total quasar
luminosity density is well-constrained at $z=2.15\pm0.05$.  We provide a
number of fitting functions to the bolometric QLF and its
manifestations in various bands, and a script$^{3}$ to 
return the QLF at arbitrary frequency and redshift from 
these fits.

\end{abstract}

\keywords{quasars: general --- galaxies: active --- 
galaxies: evolution --- galaxies: luminosity function --- cosmology: observations --- 
X-rays: galaxies --- infrared: galaxies --- ultraviolet: galaxies}

\section{Introduction}
\label{sec:intro}

Determining the nature and cosmological evolution of the quasar
luminosity function (QLF) has been of interest since quasars were
first identified as cosmological sources \citep{Schmidt68}, and
understanding the QLF is crucial to inferring 
the formation history of supermassive black holes, as
well as the buildup of cosmic X-ray and infrared (IR) backgrounds and the
contribution of quasars to reionization.  Furthermore, the recognition
that black holes appear to reside at the centers of most galaxies
\citep[e.g.,][]{KR95} and that the masses of these black holes are
correlated with either the mass \citep{Magorrian98} or the velocity
dispersion \citep{FM00,Gebhardt00} of their host spheroids,
demonstrates a link between the origin of galaxies and supermassive
black holes.  Hence, determining the evolution of the QLF is also
critical to understanding galaxy formation and evolution.

The study of the QLF has a long history
\citep[e.g.,][]{SchmidtGreen83,KooKron88,Boyle88,Hewett93,HS90,WHO94,SSG95,KDC95,Pei95},
but in recent years, surveys such as the Two Degree Field (2dF)
QSO Redshift Survey \citep[2QZ;][]{Boyle00} and the Sloan Digital Sky
Survey \citep[SDSS;][]{York00} have provided large, homogeneous quasar
samples over the range of redshifts $z=0-6$
\citep[e.g.,][]{Boyle00,Fan01a,Fan04,Croom04,2SLAQ,DR3,Jiang06a}.  In addition,
a great deal of information on the X-ray and infrared properties of
quasars has become available, and surveys with e.g.\ {\em Chandra},
{\em XMM}, {\em ROSAT}, and {\em Spitzer} have enabled studies of the
evolution of the QLF across many frequencies
\citep[e.g.,][]{Miyaji00,Ueda03,Haas04,Barger05,HMS05,Brown06,Matute06}.

Surprising and suggestive trends have emerged from these
studies.  For example, both the spectral shapes
\citep[e.g.,][]{Wilkes94,Green95,
VBS03,Strateva05,Richards06,Steffen06} and column density
distributions \citep[e.g.,][]{HGD96,SRL99,Willott00,SR00,
Steffen03,Ueda03,GRW04,Hasinger04,SazRev04,Barger05,Simpson05,Hao05}
of quasars appear to depend systematically on luminosity, with the
brightest quasars being the least obscured and the most dominated by the
optical/UV portion of the spectrum.  X-ray studies
\citep[e.g.,][]{Page97,Miyaji00,Miyaji01,LaFranca02,Cowie03,
Ueda03,Fiore03,Barger03b,HMS05} and some optical, radio, and IR
measurements \citep[e.g.,][]{Hunt04,Cirasuolo05,Matute06} indicate
that the space density of low luminosity active galactic nuclei (AGN)
peaks at redshifts lower than that of bright quasars,
specifically observing a flattening of the QLF faint-end slope with redshift. It has
been argued that this follows a similar pattern of ``cosmic
downsizing'' as has recently been observed in galaxy spheroid
populations \citep[e.g.,][]{Cowie96}.  Similarly, the bright-end slope
of the QLF appears to become shallower towards higher redshifts, from
both direct measurements \citep{Fan01b,Fan03,DR3} and
(albeit weaker) constraints from gravitational lensing
\citep{Comerford02,WyitheLoeb02,Wyithe04,Richards06lenses}.

Various models have been proposed to explain these trends, many of
which postulate that feedback from black hole growth plays a key role
in determining the black hole-host galaxy (e.g.\ $M_{\rm BH}-\sigma$)
relationships \citep{SR98,DSH05}, and co-evolution of black holes and
their host spheroids. The evolution of the faint and bright-end slopes
may be linked to these processes, with AGN feedback
providing the mechanism for cosmic downsizing, shutting down the
growth of the most massive systems at high redshift
\citep[e.g.,][]{Merloni04,Granato04,H06b,Croton06} and potentially steepening the
low-redshift bright-end QLF slope as a result \citep{Scannapieco04},
while this feedback-driven quasar decay determines the shape of the
faint-end QLF \citep{H06a}.  Feedback may also explain trends in
obscuration with luminosity, either through dust sublimation
\citep[e.g.,][]{Lawrence91} or expulsion of gas and dust on galactic
scales \citep[e.g.,][]{H05d}.  But attempts to quantitatively link the
downsizing of quasar and galaxy populations, both theoretically
\citep{H06b} and observationally \citep[e.g.,][]{H06d}, depend on a reliable
determination of the QLF.

However, inferences drawn from the observed trends suffer from
complications arising from various biases.  For example,
optical quasar surveys generally probe large volumes
($\sim10,000\,{\rm deg^{2}}$), enabling uniform sample selection at
many redshifts and the discovery of the brightest quasars, but miss
substantial populations of obscured 
\citep[e.g.,][]{RMS99,Ueda03,Treister04} or heavily reddened
\citep[e.g.,][]{Webster95,Richards01,Brotherton01,Gregg02,H04} 
quasars, and generally cover a relatively small baseline ($\sim1-2$\,dex) in
luminosity. Several seminal efforts (upon which 
we seek to expand) have been made to extend these baselines by 
compiling various optical quasar observations 
\citep[e.g.,][]{HS90,WHO94}, most notably \citet{Pei95}, but these
have still been 
severely limited by obscuration/reddening and span only $\sim2-2.5\,$dex 
in luminosity, and furthermore the 
area and depth of quasar surveys have since increased by 
orders of magnitude. 
Hard X-ray and IR samples provide a more complete census
of obscured quasars, and sample much fainter luminosities and wider
($\sim4-5$\,dex) luminosity ranges, but are measured from much smaller
survey areas ($\lesssim1\,{\rm deg^{2}}$).  Without carefully
accounting for the differential effects of obscuration, different
spectral shapes, and selection effects across bands, it is not clear
if trends observed primarily at a single frequency are physically
meaningful or robust across frequencies.  Furthermore, theoretical
models typically do not deal directly with the quasar luminosity in a
given band, but instead treat the bolometric quasar luminosity, and it
is not clear that a simple bolometric correction can reliably
translate between the two --- the effects above may change the observed
QLF shape as a function of frequency, luminosity, and redshift.

In this paper, we combine recent measurements of the QLF in many
wavelengths from the mid-IR through hard X-rays, to determine the
observed bolometric quasar luminosity function.  By utilizing
multiwavelength measurements of quasar SEDs
\citep{Elvis94,VandenBerk01,Telfer02,VBS03,Fan03,Hatziminaoglou05,
Richards06,Shemmer06} that probe the distribution of obscuration up to
and including Compton-thick column densities
\citep{RMS99,Ueda03,Treister04,Mainieri05,Hao05,Tozzi06}, we can
predict the QLFs that would be observed as a function of luminosity,
frequency, and redshift from a given bolometric QLF and compare these
simultaneously with all observed QLFs. This makes it possible to
constrain the shape and evolution of the QLF over $\sim8$ orders of
magnitude in luminosity and $\sim9$ in space density, from $z=0-6$,
larger than the baselines probed by any individual survey.

In \S~\ref{sec:data} we describe the observational data sets,
including the observed column density distributions (\S~\ref{sec:NH})
and SEDs adopted (\S~\ref{sec:bol.corr}), and consider in detail the
consistency of the QLF across various frequencies given different
simplifications of these distributions (\S~\ref{sec:tests}). In
\S~\ref{sec:bol.qlf} we calculate the bolometric QLF as a function of
luminosity and redshift, and consider the detailed evolution of the
QLF shape and its manifestation in different observed bands. In
\S~\ref{sec:lifetime} we consider a number of complementary
constraints, testing models of quasar lifetimes, the evolution in the
QLF shape, and the buildup of the black hole population
(\S~\ref{sec:models}), and we calculate the relic black hole mass
function, cosmic X-ray background (\S~\ref{sec:integrate}), and
ionization rates (\S~\ref{sec:UV}) expected from the bolometric QLF
and compare these to observations.  In \S~\ref{sec:discuss} we
summarize and discuss our results and future prospects for improving
the measurements.

We adopt a $\Omega_{\rm M}=0.3$, $\Omega_{\Lambda}=0.7$,
$H_{0}=70\,{\rm km\,s^{-1}\,Mpc^{-1}}$ cosmology, 
consistent with \citet{Spergel03,Spergel06}. 
Quasar $B$-magnitudes are Vega, and $L_{\sun}$ refers to the bolometric solar
luminosity $L_{\sun}\equiv3.9\times10^{33}\,{\rm erg\,s^{-1}}$.

\section{The Observational Data Set}
\label{sec:data}

In what follows, we compile a large number of binned QLF measurements
in different redshifts ranges, observed wavebands, and luminosity intervals.
Table~\ref{tbl:qlfs} lists the samples used, with the survey and
fields for each, the rest wavelength or band measured, the redshift
and luminosity range of the observed quasars, and the number of
quasars in the sample.
We use the 
term ``quasar'' rather loosely throughout, as the traditional $M_{B}<-23$ cut 
is not readily applicable to multiwavelength observations or obscured 
sources. Our compilation instead attempts to represent all 
AGN with {\em intrinsic} (obscuration-corrected)
luminosities above the observational limits at each redshift. This generally 
extends to the typically adopted $\sim10^{42}\,{\rm erg\,s^{-1}}$ X-ray 
luminosity limit ($L_{\rm bol}\sim10^{10}\,L_{\sun}$), below which 
confusion with normal star-forming and starburst galaxies becomes problematic
\citep[although resolved hosts at low redshift extend this to $\sim10^{9}\,L_{\sun}$; e.g.,][]{Hao05}.

Some of the observational estimates are derived from the same sets of
observations, or different updates of the same data sets.  To avoid
``double-counting,'' at the few points where two binned QLFs overlap
in both luminosity and redshift and are derived from the same sample
(at that luminosity and redshift), we discard the less
well-constrained (usually less recent) point.  This is generally rare,
and has a negligible effect on our results. It does reduce
significantly the samples of \citet{Miyaji01}, for which the Type 1
QLFs are updated in \citet{HMS05}, and \citet{Croom04}, which has many
(especially faint) luminosity intervals updated in \citet{2SLAQ}.
Some measured QLFs do not appear in this compilation
\citep[e.g.,][]{Boyle00}, as updated versions from the same or
expanded samples exist \citep[e.g.,][]{Croom04}. We have re-fit 
our results with these samples \citep[as well as those of e.g.][]{Pei95,Jiang06b,
Sazonov06,Shinozaki06,Beckmann06b} and find the differences are negligible.

\subsection{The SED and Bolometric Corrections}
\label{sec:bol.corr}

We construct a model ``intrinsic'' (un-reddened) quasar SED to compare
observations in different bands.  The template spectrum initially
follows that derived in \citet{Marconi04} and consists of a broken
power law in the optical-UV with $\alpha_{\rm O}=-0.44$
($L_{\nu}\propto \nu^{\alpha_{\rm O}}$) for $1\,\mu{\rm
m}<\lambda<1300$\,\AA\ \citep{VandenBerk01}, and $\alpha_{\rm
UV}=-1.76$ from $1200-500$\,\AA\ \citep{Telfer02}, essentially the
modal spectral slopes for optically bright blue quasars 
\citep[][appropriate given that this is supposed 
to be a pre-reddened spectrum]{Richards06}. Because a number of slightly different bands have
been used in optical quasar surveys (e.g.\ $B$, $g$, $i$, 1450\,\AA;
see Table~\ref{tbl:qlfs}), we further overlay the template spectrum of
\citet{Richards06} derived from the optically blue (i.e.\ un-reddened)
subsample onto this broken power-law.  Technically, we factor out the
best-fit power-law in the given range from the \citet{Richards06}
spectrum and then multiply our template by the residuals, but this
makes little difference compared to adopting the
\citet{Richards06} spectrum over this range. We do the same with the
template spectrum of \citet{VandenBerk01} for optical wavelengths
$1\,\mu{\rm m}<\lambda<1300$\,\AA, if a detailed correction from an
observed band to a more ``standard'' band is needed, but such
corrections have generally been provided for the QLF
measurements in Table~\ref{tbl:qlfs}.

Longwards of $\lambda>1\,\mu{\rm m}$, we adopt the mean spectrum from 
\citet{Richards06}, with a typical observed IR ``bump'' from reprocessing, eventually 
truncated as a Rayleigh-Jeans tail of blackbody emission 
($\alpha=2$), and a no-obscuration zero point taken from the optically blue
(un-reddened) subsample template spectrum.  This gives a 15$\mu{\rm m}$ to
$R$-band correction similar to the typically adopted
$\log(L_{15}/L_{R})=0.23$
\citep[e.g.,][]{Elvis94,Hatziminaoglou05,Richards06,Matute06,Brown06}
for optically unobscured (Type 1) quasars. 


The X-ray spectrum beyond 0.5\,keV is determined by a power-law, with
an intrinsic photon index $\Gamma=1.8$
\citep[e.g.,][]{George98,Perola02,Tozzi06} and an exponential cutoff
at $500$\,keV. A reflection component is included following
\citet{Ueda03}, generated with the PEXRAV model \citep{MZ95} in the
XSPEC package with a reflection solid angle of $2\pi$, inclination
$\cos{(i)}=0.5$ and solar abundances. The X-ray spectrum is then
renormalized to a given $\alpha_{\rm
ox}\equiv-0.384\,\log[L_{\nu}(2500\,$\AA$)/L_{\nu}(2\,{\rm keV})]$, and
the points at $500$\,\AA\ and $50$\,\AA\ are connected with a
power-law.  The value of $\alpha_{\rm ox}$ depends on luminosity
\citep{Wilkes94,Green95,VBS03,Strateva05}, and we adopt the most
recent determination by \citet{Steffen06},
\begin{equation}
\alpha_{\rm ox}=-0.107\,\log(L_{\nu,\,2500}/[{\rm erg\,s^{-1}\,Hz^{-1}}])+1.739, 
\end{equation}
determined specifically for unobscured (Type 1) quasars. 
The above equation is derived from the least-squares bisector 
of the $L_{\nu}(2500\,$\AA$)-L_{\nu}(2\,{\rm keV})$ relation, 
since neither luminosity can be properly taken as an ``independent'' variable -- 
the difference if e.g.\ $L_{\nu}(2500\,$\AA$)$ is considered independent 
is generally small \citep[see e.g.][]{Steffen06}, but can be of importance for 
the most luminous X-ray AGN (yielding a difference 
of $\sim30\%$ in the bolometric correction at 
$L_{0.5-2\,{\rm keV}}\sim10^{47}\,{\rm erg\,s^{-1}}$).
The baseline for these observations is sufficiently large that this
relation has been determined for nearly all luminosities of interest
in the compiled QLF measurements.  Recent comparisons between large
samples of quasars selected by both optical and X-ray surveys
\citep{RisalitiElvis05} further suggests that this is an intrinsic
correlation, not driven by e.g.\ the dependence of obscuration on
luminosity, as does our comparison of bolometric QLFs derived below.
There is no evidence for a trend of $\alpha_{\rm ox}$ with redshift,
or for any other trend in spectral shape with redshift 
\citep[e.g.,][but see also Bechtold et al.\ 2003]{Elvis94,VandenBerk01,
Telfer02,VBS03,Fan03,Steffen06,Richards06,Shemmer06}, so our spectrum
depends only on luminosity. 

\begin{figure}
    \epsscale{1.1}
    \centering
    \plotone{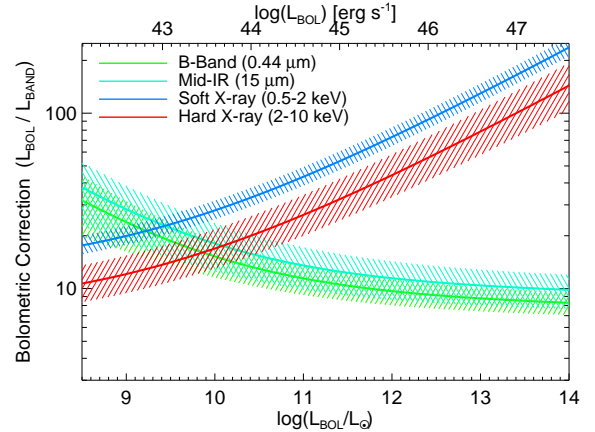}
    \caption{Bolometric corrections for $B$-band, mid-IR, soft and hard X-ray bands, 
    determined in \S~\ref{sec:bol.corr} from a number of observations as a 
    function of luminosity and given by the fitting 
    formulae in Equation~(\ref{eqn:bolcorr}). The lognormal dispersion in 
    the distribution of bolometric corrections at fixed $L$, given by 
    Equation~(\ref{eqn:bolcorr.scatter}) is shown as the shaded 
    range for each band. The full quasar 
    template spectrum is available for public download$^{3}$.
    \label{fig:bol.corr}}
    \epsscale{1.0}
\end{figure}

Ultimately, the bolometric corrections
(with zero attenuation) derived can be accurately approximated as a
double power-law
\begin{equation}
\frac{L}{L_{\rm band}}=c_{1}\,{\Bigl(}\frac{L}{10^{10}\,L_{\sun}}{\Bigr)}^{k_{1}}+
c_{2}\,{\Bigl(}\frac{L}{10^{10}\,L_{\sun}}{\Bigr)}^{k_{2}} \, ,
\label{eqn:bolcorr}
\end{equation}
with $(c_{1},\,k_{1},\,c_{2},\,k_{2})$ given by
$(6.25,\,-0.37,\,  9.00,\,-0.012)$ for $L_{\rm band}=L_{B}$,
$(7.40,\,-0.37,\,10.66,\,-0.014)$ for $L_{15\,\mu{\rm m}}$,
$(17.87,\,0.28,\,10.03,\,-0.020)$ for $L_{0.5-2\,{\rm keV}}$, and
$(10.83,\,0.28,\,  6.08,\,-0.020)$ for $L_{2-10\,{\rm keV}}$. The $k_{1}$
term is important when the given portion of the spectrum is not
dominant, controlled by the scaling of $\alpha_{\rm ox}$, and the
$k_{2}\approx0$ term represents the nearly constant bolometric
correction when a given portion of the spectrum dominates the
bolometric luminosity. Figure~\ref{fig:bol.corr} shows these corrections as a 
function of luminosity, which agree broadly with the 
values in e.g.\ \citet{Richards06} over the luminosity range they consider. 

We have generally followed \citet{Marconi04} in calculating this
spectrum, with a more detailed treatment of the optical/IR and a more
recent determination of $\alpha_{\rm ox}$, but this level of detail is
ultimately not required.  The critical dependence is that of
$\alpha_{\rm ox}$ on luminosity.  Adopting the median
\citet{Richards06} spectrum and rescaling the spectrum upwards of
$0.1\,{\rm keV}$ according to the observed $\alpha_{\rm ox}$ for
different $L_{\nu}(2500\,$\AA$)$ and corresponding bolometric
luminosities yields a similar model spectrum and essentially identical 
conclusions. Likewise, rescaling our median spectrum 
according to Equation~(21) in \citet{Marconi04} gives a
similar result.  Ignoring the dependence of bolometric
corrections on luminosity altogether, however, is problematic. In
the optical/UV, it is not as serious, as the bolometric correction
changes by only $\sim20\%$ from $M_{B}=-23$ to $-30$.  However, over a
$\sim4\,$dex interval in $L_{2-10\,{\rm keV}}$ comparable to the usual
observed baselines, the typical hard and soft X-ray bolometric
corrections change by more than an order of magnitude (primarily as
given by the $\alpha_{\rm ox}$-luminosity relation).

Finally, even accounting for the dependence of spectral shape on
luminosity, objects with a given luminosity do not all have identical
spectra and bolometric corrections. It is important to account for the
{\rm dispersion} in spectral shapes at a given $L$.  In general, there
will be two components, a correlated and uncorrelated dispersion.  If,
for example, there is a different spectral slope or value of
$\alpha_{\rm ox}$, then the bolometric correction in certain bands
will be larger, but the correction in other bands must be
smaller. Also, because the optical/UV contributes a larger fraction of
the bolometric luminosity than the X-ray, the resulting dispersion in
the $B$-band bolometric correction from different values of
$\alpha_{\rm ox}$ will be smaller than the resulting dispersion in the
X-ray bolometric corrections.

We estimate these dispersions in the power-law components of our
modeling from the observed distributions, assuming that they are
normally distributed; $\sigma_{\alpha_{\rm O}}\approx0.125$ 
\citep[][comparing the mean $\alpha_{\rm O}$ in each observed quartile]{Richards03}, 
$\sigma_{\Gamma}=0.30$ \citep{Tozzi06}, and
$\sigma_{\alpha_{{\rm ox}}}=0.075-0.14$ \citep{Steffen06}. Strictly
speaking, a larger bolometric correction in one band implies a smaller
integrated correction in others, but there are sources of scatter
which introduce this effect weakly, such as e.g.\ variations in line
strength in a given band or observational uncertainties in the
luminosity in the band. Fitting to the distribution of bolometric
corrections in \citet{Richards06}, after accounting for the luminosity
distribution of the sources and correlated dispersions above yields a
best-fit uncorrelated dispersion component $\sim0.1\,$dex, consistent
with observational estimates of the scatter in band luminosities owing
to these effects
\citep[e.g.,][]{Elvis94,George98,VandenBerk01,Perola02,Richards03,H04}.
By fitting to a number of Monte Carlo realizations of the spectra as a
function of luminosity given these dispersions, we can quantify
the effective dispersion in the bolometric correction in different
bands as a function of luminosity
\begin{equation}
\sigma_{\log{(L/L_{\rm bol})}} = \sigma_{1}\,(L_{\rm bol}/ 10^{9}\,L_{\sun})^{\beta} + \sigma_{2}
\label{eqn:bolcorr.scatter}
\end{equation}
with $(\sigma_{1},\, \beta,\, \sigma_{2}) = (0.08,\, -0.25,\, 0.060)$
in the $B$-band, $(0.07,\, -0.17,\, 0.086)$ in the IR ($15\,\mu{\rm m}$),
$(0.046,\, 0.10,\, 0.080)$ in the soft X-ray, and $(0.06,\, 0.10,\,
0.08)$ in the hard X-ray.  Here, $\sigma_{1}$ and $\beta$ roughly
describe the correlated component of the dispersion, $\sigma_{2}$ the
uncorrelated component. For typical bright quasars ($L_{\rm bol}\sim
10^{13}\,L_{\sun}$), this reflects the fact that a significant
($\gtrsim5\%$) fraction of quasars in any band can have their
bolometric luminosities mis-estimated by a factor $\sim2$ or more by a simple
bolometric correction (even one that accounts for the
luminosity-dependent spectral shape). These dispersions as a function of 
luminosity are plotted in Figure~\ref{fig:bol.corr}.

\subsection{The Observed Column Density Distribution}
\label{sec:NH}

In order to convert an observed
luminosity function to a bolometric luminosity function, we must
correct for extinction in the different observed bands, which requires the
adoption of an observed column density distribution. Essentially, 
the probability of observing a quasar of a given bolometric luminosity at 
some observed luminosity in a given band must account for the probability of 
extinction or attenuation. 

We consider three cases.  First, our fiducial model adopts the
luminosity-dependent observed column density distribution from the
hard and soft X-ray observations of \citet{Ueda03}. We also follow
\citet{Ueda03} and include an equal fraction of Compton-thick objects
with $\NH>10^{24}\,{\rm cm^{-2}}$ to that with
$\NH=10^{23}-10^{24}\,{\rm cm^{-2}}$.  The evidence for this in
\citet{Ueda03} is tentative, but it produces good agreement with the
distribution of Compton-thick column densities subsequently reported
by \citet{Treister04}, \citet{Mainieri05}, and \citet{Tozzi06} and is
consistent with upper limits to the obscured fraction from the mid-IR
observations of \citet{Richards06}. Recent very hard X-ray and 
soft gamma-ray ($\sim20-200\,$keV) {\em Swift} BAT and 
INTEGRAL observations
of local AGN, sensitive even to Compton-thick sources, confirm 
both a similar Compton-thick fraction and dependence on luminosity 
(demonstrating also that this trend does not owe to selection effects)
\citep[][but see also Wang \&\ Jiang 2006]{Markwardt05,Beckmann06a,Beckmann06b,Bassani06}. 
Note that this yields a maximum
Compton-thick fraction of $\sim30\%$ (correcting the observed number
density in a sample not sensitive to Compton-thick objects by a factor
of $1.4$), at low luminosity. This does {\em not} imply a uniform
factor of $1.4$ correction to the quasar space density, as the
Compton-thick fraction depends on luminosity in the same manner as the
entire column density distribution, and the fraction of Compton-thick
sources at high luminosity will in general be much smaller.

Alternatively, we consider a constant (luminosity-independent) column
density distribution, again adopting that from \citet{Ueda03} (fitted
to their observations assuming no luminosity dependence). As the
opposite extreme, we employ the column density distribution
determined in \citet{LaFranca05}, which depends on both luminosity and
redshift. We discuss these models in \S~\ref{sec:tests}, but find that
they are unable to produce a self-consistent set of luminosity
functions in the different observed bands.

Given an $\NH$ distribution, we calculate the extinction at X-ray
frequencies using the photoelectric absorption cross sections of
\citet{MM83} and non-relativistic Compton scattering cross sections.
In the optical and mid-IR, we adopt a canonical gas-to-dust ratio
$(A_{B}/\nh)_{\rm MW}=8.47\times10^{-22}\,{\rm cm^{2}}$ and Small
Magellanic Cloud-like reddening curve from \citet{Pei92}, which
observations suggest is appropriate for the majority of reddened
quasars \citep{Richards03,H04,Ellison05}, although a Milky-Way like
reddening curve changes the optical depth by only $\sim5-10\%$ at the
wavelengths of interest (excluding the 2100\,\AA\ ``bump'').

Given a bolometric QLF and the observed column density 
distribution, we can then convolve over the distribution of
column densities and spectral shapes at each bolometric luminosity $L$ to infer the implied
distribution of luminosities that should be observed in a given
band. In other words, knowing the probability of some intrinsic spectral shape 
and intervening column density, we determine the probability of observing 
quasars with an intrinsic $L$ at some observed luminosity in the observed band(s). 

It is a surprisingly good approximation to this full convolution
to adopt an ``obscured fraction'' as a function of luminosity; i.e.\
converting the luminosities of the 
bolometric QLF to luminosities in the observed band using the 
appropriate bolometric corrections 
and then multiplying by some ``observable fraction'' $f(L)$ 
to correct for the effects of extinction (and the dispersion in spectral shapes). 
Fitting to this $f(L)$ from our full modeling yields a
useful function in comparing optical, soft X-ray, and hard X-ray
observations; more directly applicable than the
typically calculated fraction with
$\NH>10^{22}\,{\rm cm^{-2}}$. This can be conveniently parameterized as a power law, 
\begin{equation}
f(L)\equiv\frac{\phi(L_{i})}{\phi(L[L_{i}])}=f_{46}\,
{\Bigl(}\frac{L}{10^{46}\,{\rm erg\,s^{-1}}}{\Bigr)}^{\beta} \, ,
\label{eqn:obsc}
\end{equation}
where $L_{i}$ is the luminosity in some band, and $L$ is the corresponding 
bolometric luminosity given by the bolometric corrections of Equation~(\ref{eqn:bolcorr}). 
This gives values $(f_{46},\,\beta)$ of $(0.260,\,0.082)$ for $L_{i}=L_{B}$ ($4400\,$\AA), 
$(0.438,\, 0.068)$ for $L_{i}=L_{IR}$ ($15\,\mu{\rm m}$), $(0.609,\,0.063)$ for 
$L_{i}=L_{SX}$ ($0.5-2\,{\rm keV}$), 
and $(1.243,\,0.066)$ for $L_{i}=L_{HX}$  ($2-10\,{\rm keV}$). 
Note that this ``observable fraction'' 
can exceed unity, because the scatter and luminosity 
dependence of the bolometric corrections significantly changes the 
shape of the bright-end QLF in the X-ray bands (see Figure~\ref{fig:show.demo} below). 
These simple rescalings are robust for 
the $B$ and IR bands, with weak dependence on the shape of the bolometric QLF and 
dispersion in bolometric corrections, but the effects above make the observable 
fraction in the X-ray bands sensitive to the shape of the QLF; if greater 
accuracy is required, a more robust fit across a variety of QLF shapes gives 
$\beta=0.035\slopebright,\ 0.034\slopebright$ for $L_{SX}$ and $L_{HX}$, respectively, 
for an arbitrary bright-end QLF slope $\slopebright$ (defined in \S~\ref{sec:fits.at.z}). 
We caution that these are crude approximations, but the above 
equations can be used for rough conversion of observed QLFs to 
bolometric QLFs and vice versa, and for average conversions between bands 
(multiplying any two such conversions together appropriately).

\subsection{Comparison of the Bolometric QLF Under Different Assumptions}
\label{sec:tests}

Given some model for the quasar spectrum and column density
distribution, we can calculate the bolometric QLF from observations
in some bands.  In detail, we convolve
a given bolometric QLF over the distribution of intrinsic spectral
shapes and column densities, which yields the expected 
luminosity distribution in some observed band.  By integrating, if
necessary, over the appropriate redshift and luminosity intervals, we
can directly compare this to the binned QLF measurements at each frequency,
luminosity, and redshift. We minimize the $\chi^{2}$ of this
estimate in relation to the observed QLF in all bands, 
\begin{equation}
\chi^{2}\equiv\sum{\Bigl(}\frac{\phi_{\rm expected}(L_{\nu},\,z\,|\,\phi_{\rm bol}) - \phi_{\rm obs}(L_{\nu},\,z)}
{\Delta\phi_{\rm obs}(L_{\nu},\,z)}{\Bigr)}^{2}\,,
\end{equation}
to determine the
best-fit bolometric QLF.  Before applying this 
broadly, we would to test our
description of the observed spectral shape and column density
distributions. 

\begin{figure*}
    \centering
    \plotone{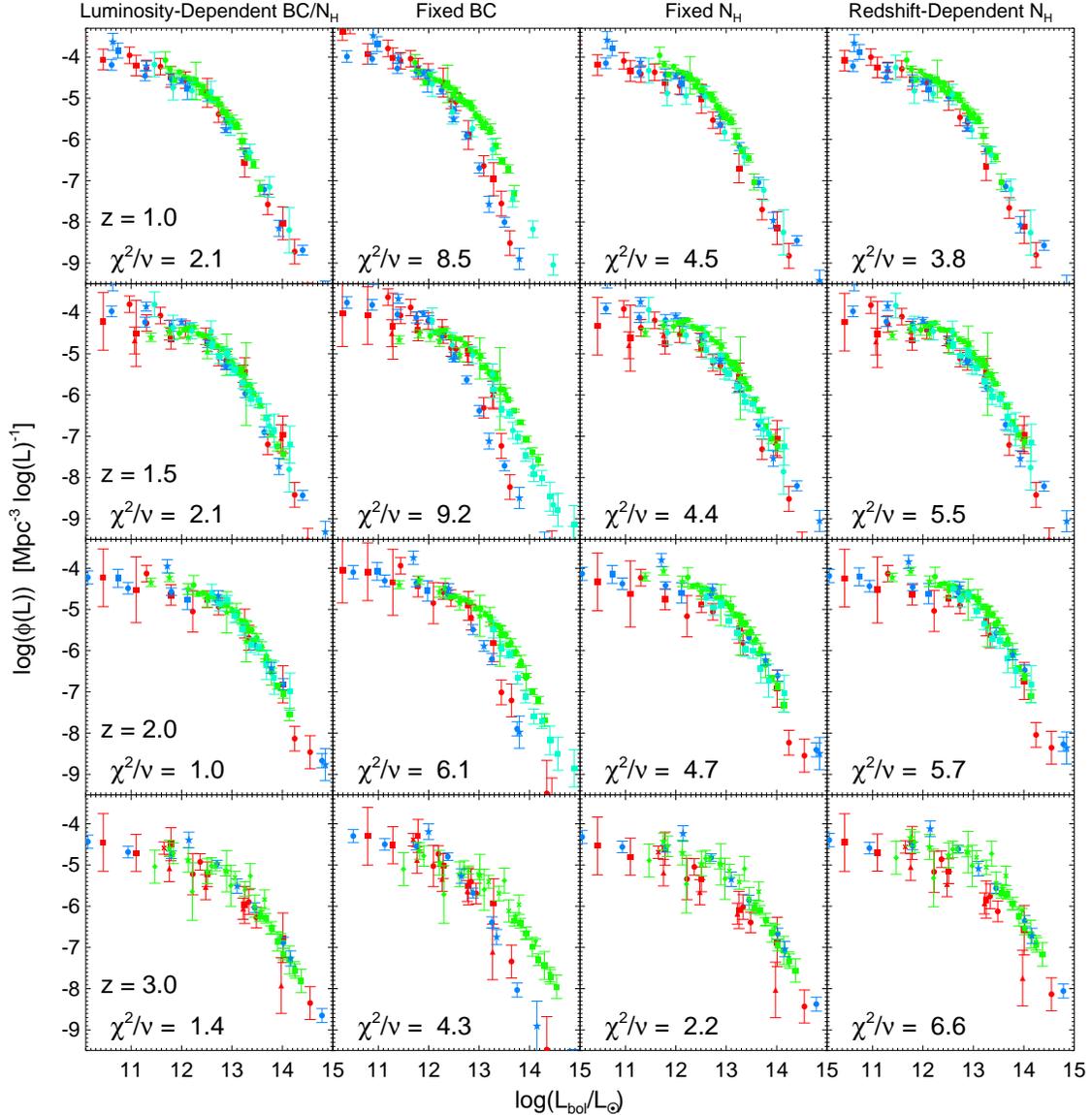}
    \caption{The bolometric QLF from the 
    observations in Table~\ref{tbl:qlfs} at various
    wavebands (optical: green, 
    soft X-ray: blue, hard X-ray: red, IR: cyan; see 
    Table~\ref{tbl:qlfs} for the plotting symbols).  For ease of
    comparison, the points in the different bands are rescaled 
    as described in the text to indicate the bolometric QLF implied
    by each set of measurements.
    Left panels 
    show the bolometric QLF adopting our full (luminosity-dependent) bolometric 
    corrections and column density distributions \citep[see also][]{Ueda03,Marconi04,LaFranca05},  
    at several redshifts (as labeled). Center-left shows the result with the full column 
    distribution, but adopting the constant bolometric corrections of \citet{Elvis94}. 
    Center-right uses the full bolometric corrections, but a luminosity-independent 
    column density distribution (obscured fraction). Right uses the full 
    bolometric corrections and a strongly redshift-dependent column density distribution 
    from \citet{LaFranca05}. Each panel shows the reduced $\chi^{2}$ for 
    the assumption that the data yield a consistent bolometric QLF at each redshift. 
    The latter three assumptions do not yield a consistent bolometric QLF at each redshift, 
    unlike for the observationally derived luminosity-dependent bolometric 
    corrections and column density distributions. 
    \label{fig:binned.bol.compare}}
\end{figure*}

In Figure~\ref{fig:binned.bol.compare}, we show (left panels) the
bolometric QLF determined in this manner at several redshifts 
from our full modeling.  At each redshift, we compare
with QLF observations from Table~\ref{tbl:qlfs} with overlapping
redshift intervals.  We plot the binned QLF measurements rescaled to
bolometric points with respect to the best-fit double power law
bolometric QLF at that redshift (see \S~\ref{sec:bol.qlf} below).  In
other words, convolving our best fit bolometric QLF with e.g. the
quasar spectrum and column density distribution predicts a number
density of quasars $n_{\rm mdl}$ for a given observed bin.  Comparing
this to the actual number observed, $n_{\rm obs}$, fixes the ratio
$n_{\rm obs}/n_{\rm mdl}$.  In Figure~\ref{fig:binned.bol.compare} we
show the agreement with all bands simultaneously by plotting $n_{\rm
obs}/n_{\rm mdl}$ times the best-fit bolometric QLF at each luminosity
as the colored points.  
Given our full SED and obscuration model, 
the inferred bolometric QLF from each observed waveband agrees well.  We quantify this
directly, showing in each panel the $\chi^{2}/\nu$ statistic
corresponding to the probability that all of the observations from the
different bands derive from a single (technically double power-law,
although this assumption only weakly changes the $\chi^{2}/\nu$)
bolometric QLF. 

\begin{figure*}
    \centering
    \plotone{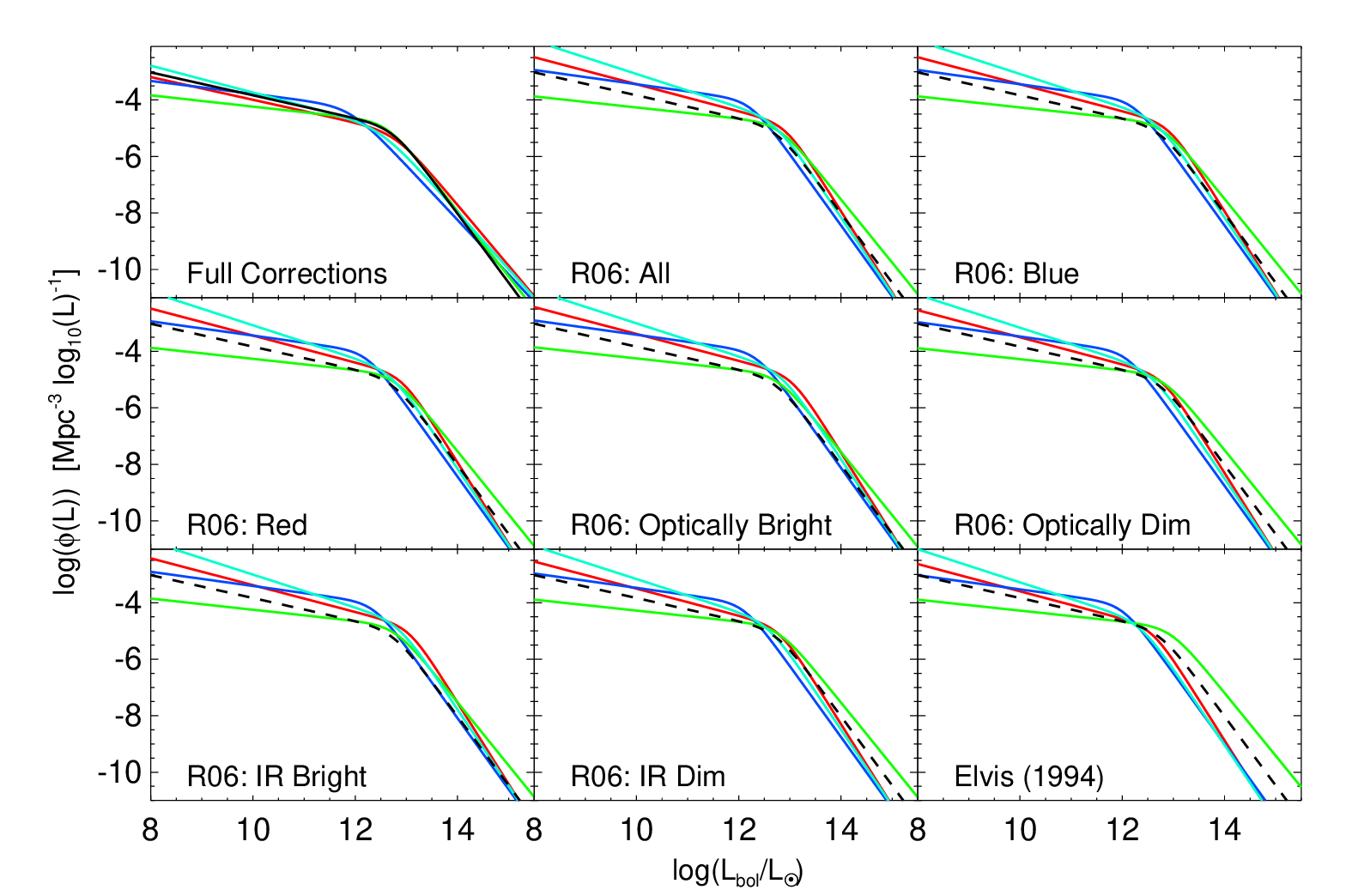}
    \caption{Comparison of the best-fit bolometric QLFs to the observations 
    in Figure~\ref{fig:binned.bol.compare} at $z\sim1$, 
    fitted independently to the optical (green), soft X-ray 
    (blue), hard X-ray (red), and IR (cyan) data sets. Upper left panel shows 
    the result using our full (luminosity-dependent) bolometric corrections, 
    with the black line the bolometric QLF from fitting to the data at all wavelengths 
    simultaneously (reproduced as the dashed black line in the other panels), 
    subsequent panels adopt constant (luminosity-independent) 
    corrections from \citet{Richards06} based on the mean spectrum of the 
    complete (all), blue, red, optically bright, optically dim, IR bright, and IR dim 
    sub-samples considered therein. Lower right panel adopts the 
    \citet{Elvis94} mean spectrum. 
    Accounting for the dependence of spectral shape on luminosity 
    yields a consistent bolometric QLF across all frequencies and luminosities, 
    whereas considering 
    a single mean spectrum is generally appropriate only for narrow 
    luminosity and color intervals. 
    \label{fig:const.corr.compare}}
\end{figure*}

If, however, we simplify our modeling of observed quasar spectra or
column densities, the consistency between the QLF implied from the
different wavebands is broken. First, we consider the prediction if
all quasars had a single spectral shape by adopting the model spectrum
of \citet{Elvis94}, with no dependence on luminosity. The optical, IR,
and X-ray observations are no longer consistent with one another, and
the $\chi^{2}/\nu$ at each redshift shown rises by a factor $\sim4$ to
an unacceptably high value. Hard and soft X-ray measurements
are still consistent, as they are in both cases 
related by a relatively straightforward power law.
These general conclusions are unchanged regardless of the exact 
quasar spectrum adopted. In \citet{Richards06}, 
mean spectra are computed for
the entire quasar sample as well as for several sub-samples: the most
optically luminous/dim, optically red/blue, and IR luminous/dim halves
(divided at the median values $\log{[L_{\rm opt}/({\rm
erg\,s^{-1}})]}=46.02$, $\Delta(g-i)=-0.04$, and $\log{[L_{\rm
IR}/({\rm erg\,s^{-1}})]}=46.04$).  Figure~\ref{fig:const.corr.compare} compares 
the bolometric QLF as
in Figure~\ref{fig:binned.bol.compare}, at $z=1$. For clarity, we
show only the best-fit double power-law rather than the binned
points. Considering the 
different \citet{Richards06} and \citet{Elvis94} mean spectra, 
no single quasar spectrum yields a consistent bolometric QLF at all
luminosities (i.e.\ there is no ``effective mean'' spectral shape). 
Taking the mean spectrum of the most optically or IR
bright quasars, unsurprisingly, yields a consistent bolometric QLF at
the bright end (above the break).  The optically and IR dim spectra,
on the other hand, are appropriate for lower-luminosity quasars (those
near and just below the break), and produce consistent bolometric QLFs
in this regime.  Note that this does {\em not} extend to the lowest
luminosities shown, as the ``optically dim'' objects in
\citet{Richards06} are still much brighter ($L_{\rm bol}\gtrsim 10^{45}\,{\rm
erg\,s^{-1}} \approx 3\times10^{11}\,L_{\sun}$)
than the lowest luminosities plotted in Figure~\ref{fig:const.corr.compare} and probed
by X-ray samples.

We next examine different assumptions for the column density
distribution. In Figure~\ref{fig:binned.bol.compare}, 
we adopt the full spectral model but consider the 
luminosity-independent mean column density distribution 
from \citet{Ueda03}; i.e.\ a constant ratio of
obscured to unobscured AGN $\sim 2:1$, similar to that found locally
\citep[e.g.,][]{RMS99,Hao05}. 
Again, the agreement is broken: 
quasars at low luminosities, where the 
number of quasars in the \citet{Ueda03} sample is large, dominate the
mean $\NH$ distribution, and as a result this constant obscured fraction provides an
acceptable (although still less good) fit to the low-luminosity data.
However, at high luminosities, the results derived from 
different wavebands diverge. The total $\chi^{2}/\nu$
unambiguously rises (factor $\sim2$) 
at all redshifts, with an increasing discrepancy at higher luminosities 
(factor $\sim3$ increase in $\chi^{2}/\nu$ above the QLF break). The
disagreement between different bands changes significantly as a
function of luminosity, so re-normalizing the constant
obscured fraction cannot make the luminosity functions consistent.

Finally, we adopt a column density distribution which evolves strongly
with redshift (and luminosity).  Specifically, we adopt the fit from
\citet{LaFranca05} to the $\NH$ distribution with maximal redshift
evolution, similar to the redshift evolution implied by the column
density distributions in \citet{Tozzi06} (fitted assuming maximal
redshift dependence and no luminosity dependence), and similar to the
redshift evolution of obscured fractions implied by some X-ray
background synthesis models (e.g., Comastri et al.\ 1995; Gilli et
al.\ 1999; but see also Treister \& Urry 2005). The resulting
bolometric QLF is shown in the right panels of
Figure~\ref{fig:binned.bol.compare}.

This fit is reasonable at low and moderate redshifts $z\lesssim1$ 
(unsurprisingly, where the observed samples are best constrained). 
Extrapolated to higher redshifts, the
consistency between different bands breaks down,
increasing $\chi^{2}/\nu$ by an unacceptable factor $\sim3$ by
$z\gtrsim2$.  Therefore, while our analysis implies that the
obscured fraction must not be constant, it is also clear that any
variation is primarily a luminosity, and not a redshift
dependence. However, more moderate redshift evolution such as that
suggested by \citet{Barger05} or \citet{Ballantyne06}, with obscured
fractions evolving by only $\sim20\%$ from $z=0-1$ (with no further
evolution at higher redshifts), is consistent with our analysis, and
will have only a small impact on the cumulative QLF evolution.  Such
evolution, being a small effect, would be better constrained by direct
measurements of the column density distribution at moderate redshifts.

Likewise, although not shown in Figure~\ref{fig:binned.bol.compare},
we find no evidence for any dependence of bolometric corrections on
redshift, in agreement with a number of direct measurements of quasar
SEDs \citep[e.g.,][]{Elvis94,VandenBerk01,
Telfer02,VBS03,Fan03,Steffen06,Richards06,Shemmer06}.  If the
variation of e.g.\ $\alpha_{\rm ox}$ were primarily a redshift as
opposed to luminosity dependence, or if there were e.g.\ a strong
dependence of X-ray photon index on redshift as suggested by
\citet{Kuhn01} and \citet{Bechtold03}, then our inferred bolometric QLFs from
different bands would diverge with redshift.  Instead, they appear to
be self-consistent given the redshift-independent spectral model at
all $z=0-6$. Given the limitations of present data, however, it is still possible 
that less well-constrained portions of the spectrum evolve differently. For example,  
the IR SED could, in principle, evolve differently with luminosity or redshift 
from the optical SED \citep[but see][who find similar near-IR SEDs at high redshift]{Jiang06b}, 
and \citet{Maiolino04} have suggested that extinction curves may evolve 
at $z\gtrsim4-5$ (although their proposed extinction curves change the 
optical depth at the wavelengths of interest by $\lesssim10\%$). There is no 
strong evidence for such additional evolution in the samples we consider, 
but these caveats should be considered in any extrapolation of 
our fitting to less well-sampled frequencies and redshifts.

\subsection{The Relation of Bolometric to Observed QLFs}
\label{sec:bol.to.obs}

\begin{figure*}
    \centering
    \plotone{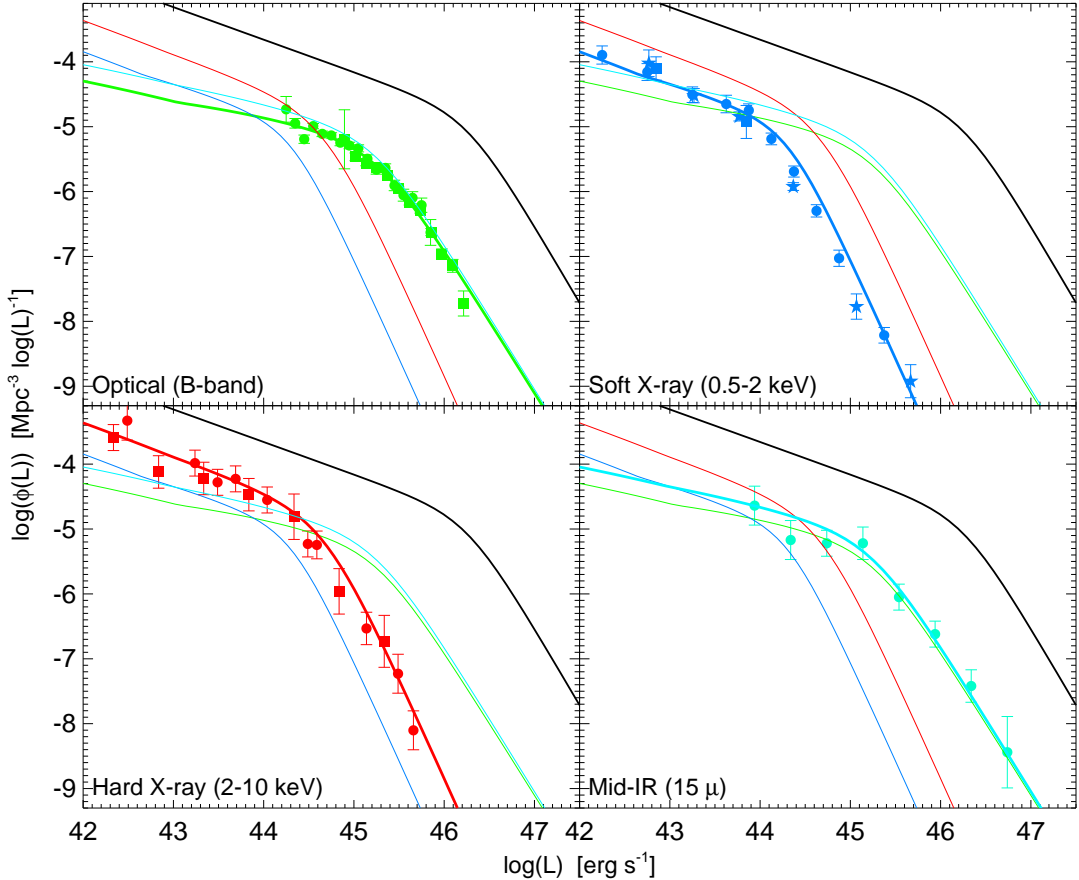}
    \caption{The best-fit bolometric QLF at $z\sim1$ (black lines), with the 
    resulting observed QLF in each of several bands: optical (green), 
    soft X-ray (blue), hard X-ray (red), and IR (cyan). Each corresponding 
    panel plots the compiled observations in Table~\ref{tbl:qlfs} 
    for the appropriate redshift and frequency (points). Symbols denote the 
    parent sample as given in Table~\ref{tbl:qlfs}. The observations at each 
    band are consistently produced from a single bolometric QLF, and together 
    provide strong constraints on the QLF shape. Different wavebands accurately 
    represent different aspects of the bolometric QLF shape. With only a weakly 
    $L$-dependent bolometric correction but significant effects of obscuration, 
    the optical QLF faithfully traces the bolometric bright end shape, but 
    is flatter at faint $L$. Hard X-rays, 
    conversely, are weakly affected by obscuration but can have a strongly 
    luminosity-dependent bolometric correction, reproducing the faint end shape 
    but being steeper at bright $L$. The IR QLF better follows the bolometric shape,
    but is currently least well-constrained. 
    \label{fig:show.demo}}
\end{figure*}

We briefly examine the relation between the shape of the bolometric QLF
and the observed QLF at different frequencies.
Figure~\ref{fig:show.demo} shows the fit to the bolometric QLF from
Figure~\ref{fig:binned.bol.compare}, at $z\sim1$, and the resulting 
observed QLF in different bands. Although our
fitting considers all observations in the bands given in
Table~\ref{tbl:qlfs}, we have renormalized all optical observations to
the $B$-band for plotting purposes (likewise
for the other bands).  We plot all observations in the same units
($B$-band and $15\,\mu{\rm m}$ show $\nu L_{\nu}$) for the sake of direct
comparison.

The optical/UV spectrum dominates the bolometric luminosity (at least
in bright, unobscured quasars), and therefore the optical bolometric
correction depends only weakly on luminosity.  Consequently, the
bright-end slope of the optical QLF is essentially identical to that
of the bolometric QLF, offset by a nearly constant (when calibrated
for optically bright quasars) bolometric correction.  However, at the
faint end, obscuration significantly flattens the optical QLF. In the
hard X-ray, this is reversed. The faint-end slope is relatively
unaffected by obscuration and directly probes the bolometric
faint-end, but the change in bolometric corrections (as well as the
scatter in bolometric corrections at fixed $L$) alters the bright-end
shape considerably. The IR, unsurprisingly, provides in many ways the
most direct estimate of the bolometric luminosity distribution, but is
the most poorly constrained by current observations.  The soft X-ray,
while yielding some of the tightest observational constraints and
critical for evaluating e.g.\ the UV contributions of quasars, is
affected at the faint and bright ends by obscuration and changing
bolometric corrections, respectively.

\subsection{The Contribution of Quasar Host Galaxy Light}
\label{sec:host.lum}

We have so far ignored the contribution of quasar host galaxies 
to their observed luminosities. Observed luminosity functions, however, 
do not generally remove this host contamination. In the X-ray bands, 
this is not of course expected to be a serious concern, and in 
the emission-line luminosity functions of \citet{Hao05} we consider 
their more conservative AGN cut which should eliminate much of the 
contamination from star-forming nuclear regions in the hosts. 
However, this contribution may still be problematic at optical 
and IR wavelengths. 
If we knew the black hole masses of quasars in the observed QLFs, it would 
be straightforward to at least estimate an average host contribution or bias, 
as there is a well-established correlation between black hole mass and 
host galaxy optical luminosity (or mass) 
\citep[e.g.,][]{KR95}. Lacking this information, 
we must adopt some estimate of the observed quasar Eddington ratio ($L/L_{\rm Edd}$)
distribution to convert from a bolometric luminosity to black hole mass and 
corresponding host luminosity. 

\begin{figure}
    \epsscale{1.1}
    \centering
    \plotone{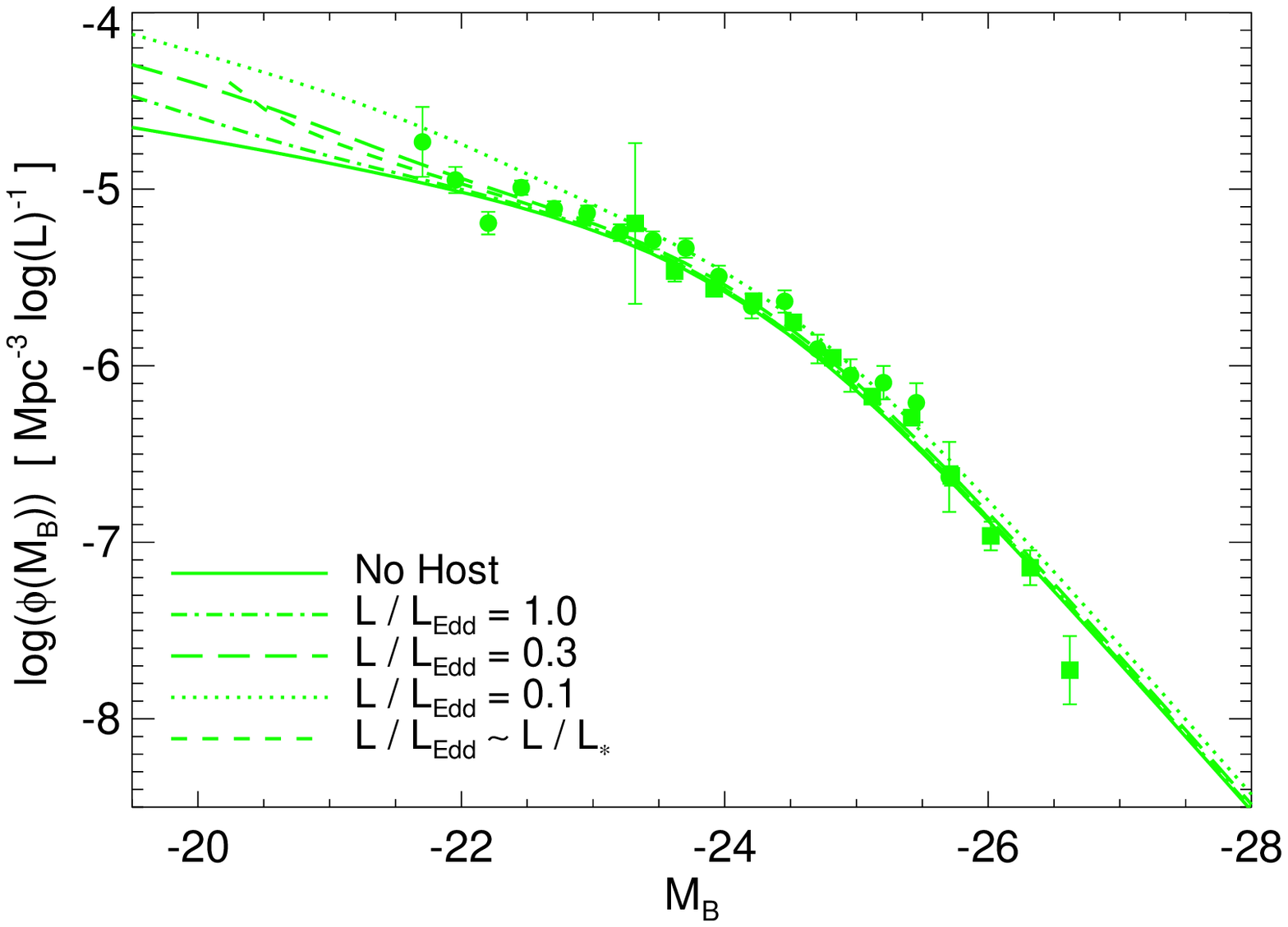}
    \caption{The observed $B$-band QLF (points) at $z\sim1$, and that determined from the 
    best-fit bolometric QLF in Figure~\ref{fig:show.demo} (solid line) 
    ignoring contributions to the observed luminosity from 
    quasar host galaxies. Also shown is the $B$-band QLF determined from the same bolometric 
    QLF, but including the expected host galaxy contribution to the observed luminosities
    (normalized for a given Eddington ratio with the observed relation between 
    black hole mass and host galaxy $B$-band luminosity from 
    \citet{MarconiHunt03}), assuming all quasars have constant 
    $L/L_{\rm Edd}=1.0,\ 0.3,\ 0.1$ (dot-dashed, long-dashed, dotted lines, respectively), or 
    assuming $L/L_{\rm Edd}=L/L_{\ast}$ (short dashed line; $L_{\ast}$ is the 
    QLF break luminosity). Host galaxy contributions should be negligible at most 
    luminosities and redshifts for the observed bands we consider, but will be 
    important for future optical, near- and mid-IR surveys which probe 
    fainter luminosities $L\lesssim10^{12}\,L_{\sun}$ ($M_{B}\gtrsim-23$). 
    \label{fig:show.host}}
    \epsscale{1.0}
\end{figure}

In Figure~\ref{fig:show.host}, we consider 
several simple, representative cases to estimate the possible effects of 
our neglecting host luminosities. 
We compare the $B$-band QLF neglecting host light (our fiducial model) 
and alternatively, assuming all quasars 
are radiating at the Eddington luminosity. This gives a black hole 
mass for each $L$, and we use the black hole mass-host $B$-band 
luminosity relationship from \citet{MarconiHunt03} to determine the corresponding 
host contribution \citep[see also][who measure a similar $L/L_{\rm Edd}$-dependent 
relation directly in AGN]{VandenBerk06}. The quasar luminosities are corrected and attenuated 
as before, and then the (un-attenuated) host luminosity is added. 

Strictly speaking, this 
is not necessarily self-consistent, as the observational calibrations of bolometric 
corrections do not necessarily remove host galaxy contributions. However, our 
construction of the intrinsic quasar SED should effectively exclude 
host contributions (compare e.g.\ \citet{Richards06}
who explicitly remove a rough lower limit to the host contribution), so long as 
there is not a substantial host contribution at $\sim2500\,$\AA\ which would 
significantly affect the determination of $\alpha_{\rm ox}$ (although this is not 
expected at these frequencies, and the relation in \citet{Steffen06} is calibrated from only 
Type 1 AGN). In any case, these effects are second-order to those shown in 
Figure~\ref{fig:show.host}.

We also consider the case where all quasars radiate 
at an Eddington ratio $L/L_{\rm Edd}=0.3$ and $L/L_{\rm Edd}=0.1$.
As an alternative simplification, we consider the case if the QLF is a pure Eddington ratio 
sequence with $L/L_{\rm Edd}=L/L_{\ast}$ ($L_{\ast}$ is the 
fitted QLF break luminosity). It makes no difference 
if we allow $L/L_{\rm Edd}>1$ or fix the maximum $L/L_{\rm Edd}=1$, 
as the host contribution is minimal in either case. We show this for 
the $B$-band, but the effects in the mid-IR are similar (given e.g.\ the typical 
early-type spectrum of \citet{FRV97}; effects are also similar for other host types). 

Figure~\ref{fig:show.host} demonstrates that including
quasar host galaxy luminosities should be a small effect, except at the 
lowest luminosities $M_{B}\gtrsim-23$ (bolometric $L\lesssim10^{12}\,L_{\sun}$). 
This is unsurprising, as at bright 
observed $M_{B}$, obscuration is small and Eddington ratios are expected to 
be high. In general, this implies a negligible impact on our subsequent calculations, 
as the deepest optical observations only just probe the range in which 
host luminosity contamination becomes significant (the larger baseline at 
low luminosities comes from hard X-ray observations, 
which do observe significant numbers of optically normal galaxies 
hosting low-luminosity X-ray AGN, e.g.\ Barger et al.\ 2005). At higher redshift, 
flux limited samples are further removed from these luminosities, and 
Eddington ratios are expected to be uniformly high \citep[at least in 
optically-selected samples, e.g.][]{MD04}. Moreover, observations of 
the relation between black hole mass and host luminosity 
at high redshift, albeit considerably uncertain, suggest that host luminosities
either remain constant or decrease at a given $M_{\rm BH}$
\citep[e.g.,][]{Peng06}. 

However, this strongly cautions the interpretation of
deeper optical, near- or mid-IR surveys, especially at low redshifts $z\lesssim0.5$, 
where Eddington ratios may well be low ($\lesssim0.1$). If such surveys 
seek to probe luminosities $\lesssim10^{12}\,L_{\sun}$, careful consideration 
of the contribution from quasar hosts is necessary to determine the 
actual quasar contribution at low luminosities. For example, 
Figure~\ref{fig:show.host} demonstrates 
that the observed faint-end QLF slope in these bands can be significantly biased 
by host contamination at magnitudes lower than those currently probed, 
in a manner sensitive to 
the Eddington ratio or black hole mass distribution. These effects will be 
even more pronounced in the near-IR, as the ratio of host to 
quasar luminosity in unobscured objects has a typical maximum at 
$\nu\sim1.6\mu {\rm m}$ \citep[see e.g.\ Figure~11 of][]{Richards06}.

\section{The Bolometric QLF}  
\label{sec:bol.qlf}
\subsection{The Bolometric QLF at Specific Redshifts}
\label{sec:fits.at.z}

We combine the binned QLF measurements to examine the QLF ``at'' a
single redshift.  Of course, the observations are not made over
identical redshift intervals, so we re-normalize each to the same
redshift (as $n_{\rm obs}/n_{\rm mdl}$) with the best-fit analytic QLF
fit from the same (or appropriate companion) paper. This introduces
some additional model dependence, so these fits should be considered
as heuristic, but they will inform our subsequent choice of functional
forms for more properly fitting to the redshift dependence of the
bolometric QLF.

We follow standard practice and fit the QLF to a double power law
\begin{equation}
\phi(L)\equiv \frac{{\rm d}\Phi}{{\rm d}\log(L)}=\frac{\phistar}
{(L/\lstar)^{\slopefaint}+(L/\lstar)^{\slopebright}}, 
\label{eqn:doublepwrlaw}
\end{equation}
with normalization $\phistar$, break luminosity $\lstar$, faint-end
slope $\slopefaint$, and bright-end slope $\slopebright$.  Note that
conventions for the double power law are often different in optical
and X-ray analyses; for optical QLFs typically the double power
law is defined in terms of \citep[e.g.,][]{Peterson97,Croom04}
\begin{equation}
\frac{{\rm d}\Phi}{{\rm d}L}= \frac{\phistar'/\lstar}
{(L/\lstar)^{-\alpha}+(L/\lstar)^{-\beta}}, 
\end{equation}
or per unit absolute magnitude 
\begin{equation}
\frac{{\rm d}\Phi}{{\rm d}M}= \frac{\phistar''}
{10^{0.4\,(\alpha+1)\,(M-M_{\ast})}+10^{0.4\,(\beta+1)\,(M-M_{\ast})}}, 
\end{equation}
which in our notation gives $\alpha=-(\slopefaint+1)$, $\beta=-(\slopebright+1)$, 
$\phistar'=\phistar/\ln{10}$, and $\phistar''=0.4\,\phistar$. 

\begin{figure*}  
    \centering
    \plotone{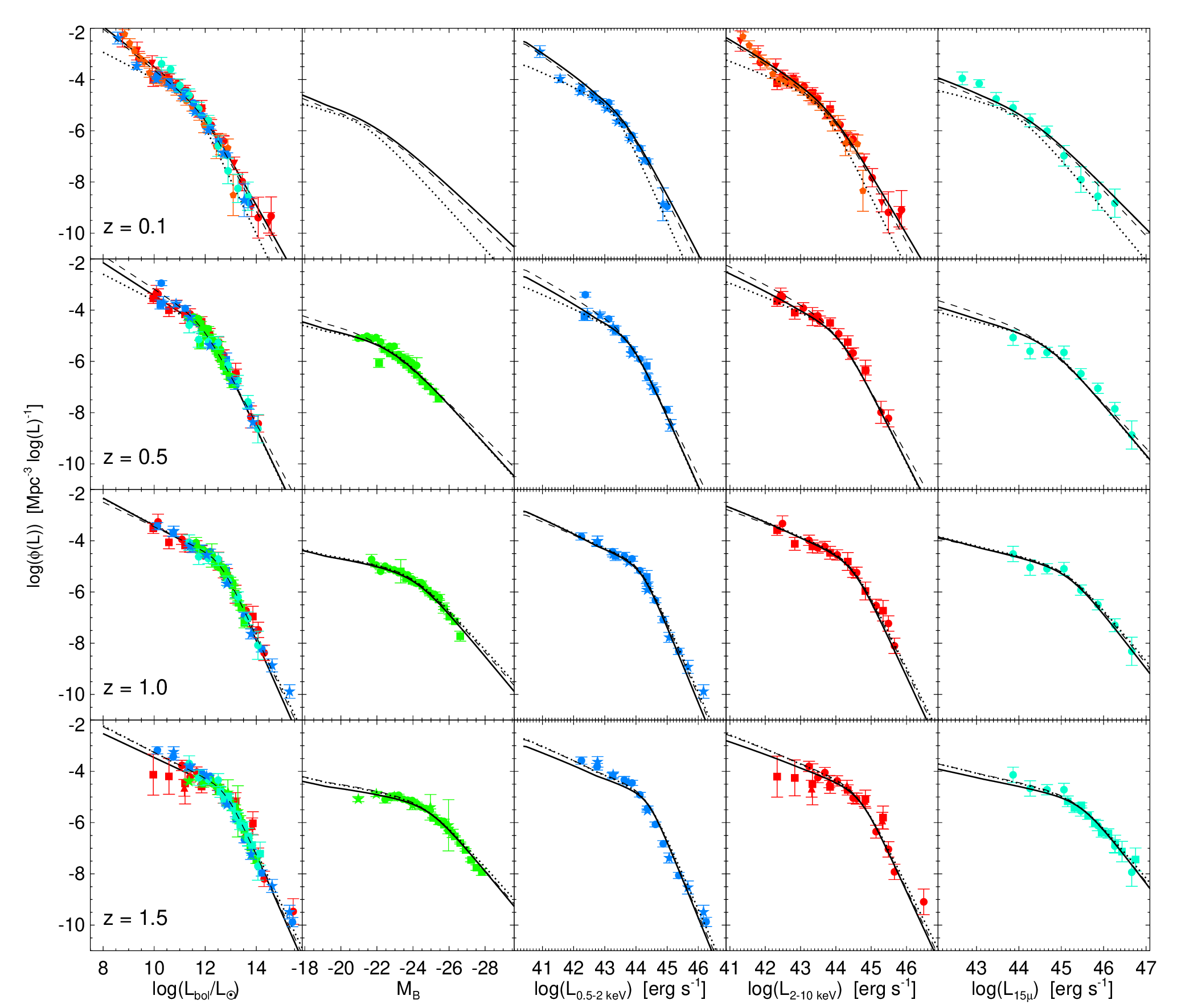}
    \caption{The best-fit bolometric QLF at each of several redshifts 
    (left panels; shown as $n_{\rm mdl}/n_{\rm obs}$), and the 
    corresponding observed QLF in $B$-band (center-left; green), 
    soft X-rays ($0.5-2$\,keV) (center; blue), 
    hard X-rays ($2-10$\,keV) (center-right; red), and 
    mid-IR ($15\,\mu$m) (right; cyan). Rather than add a series of
    panels for a single data set, the emission-line luminosity functions 
    of \citet{Hao05} are shown (orange) in the $z=0.1$ hard X-ray panel
    (rescaled by $n_{\rm obs}/n_{\rm mdl}$, but equivalently directly 
    converted to hard X-ray luminosities following Heckman et al.\ 2005).
    Lines show the best-fit 
    evolving double power-law model to all redshifts (solid), the best-fit 
    model at the given redshift (dashed), and the best-fit PLE model (dotted). 
    Points shown are the compiled observations from Table~\ref{tbl:qlfs}, 
    with the plotting symbols for each observed sample listed therein. 
    \label{fig:show.z.low}}
\end{figure*}

\begin{figure*}
    \centering
    \plotone{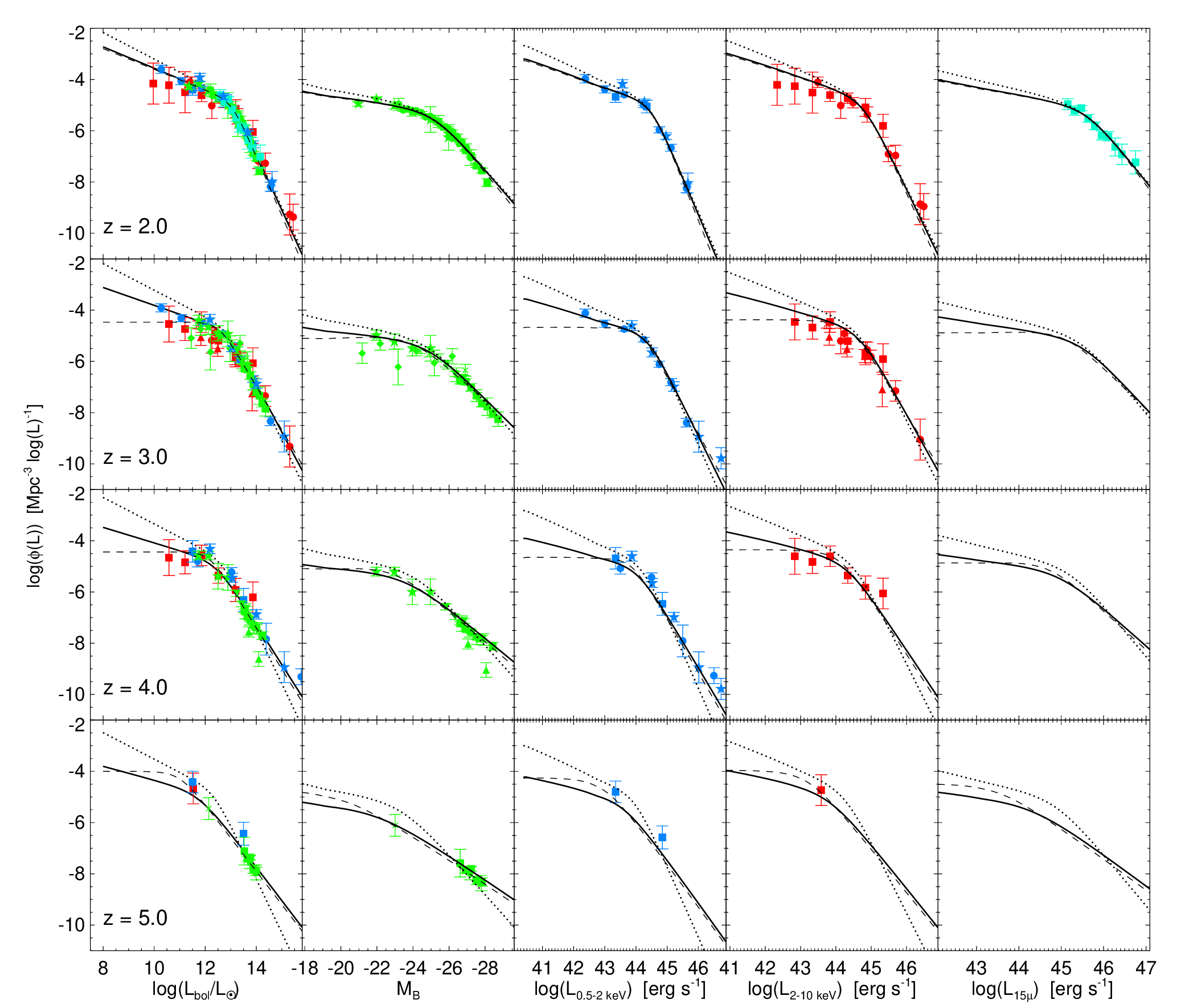}
    \caption{As Figure~\ref{fig:show.z.low}, but at higher redshifts, as labeled. 
    \label{fig:show.z.high}}
\end{figure*}

The resulting bolometric QLF at several redshifts is shown in
Figures~\ref{fig:show.z.low} \& \ref{fig:show.z.high}. At each redshift, all
binned observations from Table~\ref{tbl:qlfs} in overlapping redshift
intervals are shown. The
QLFs from optical, soft and hard X-ray, IR, and emission-line
measurements yield a self-consistent bolometric QLF at all
luminosities and redshifts probed, when a full modeling of the
observed spectral and $\NH$ distributions is employed.  Furthermore, the
combination of this number of observations provides a large
luminosity and redshift baseline, sampling the the faint end, break,
and bright-end slope to redshifts $z\sim4-5$.


\begin{figure*}
    \centering
    \plotone{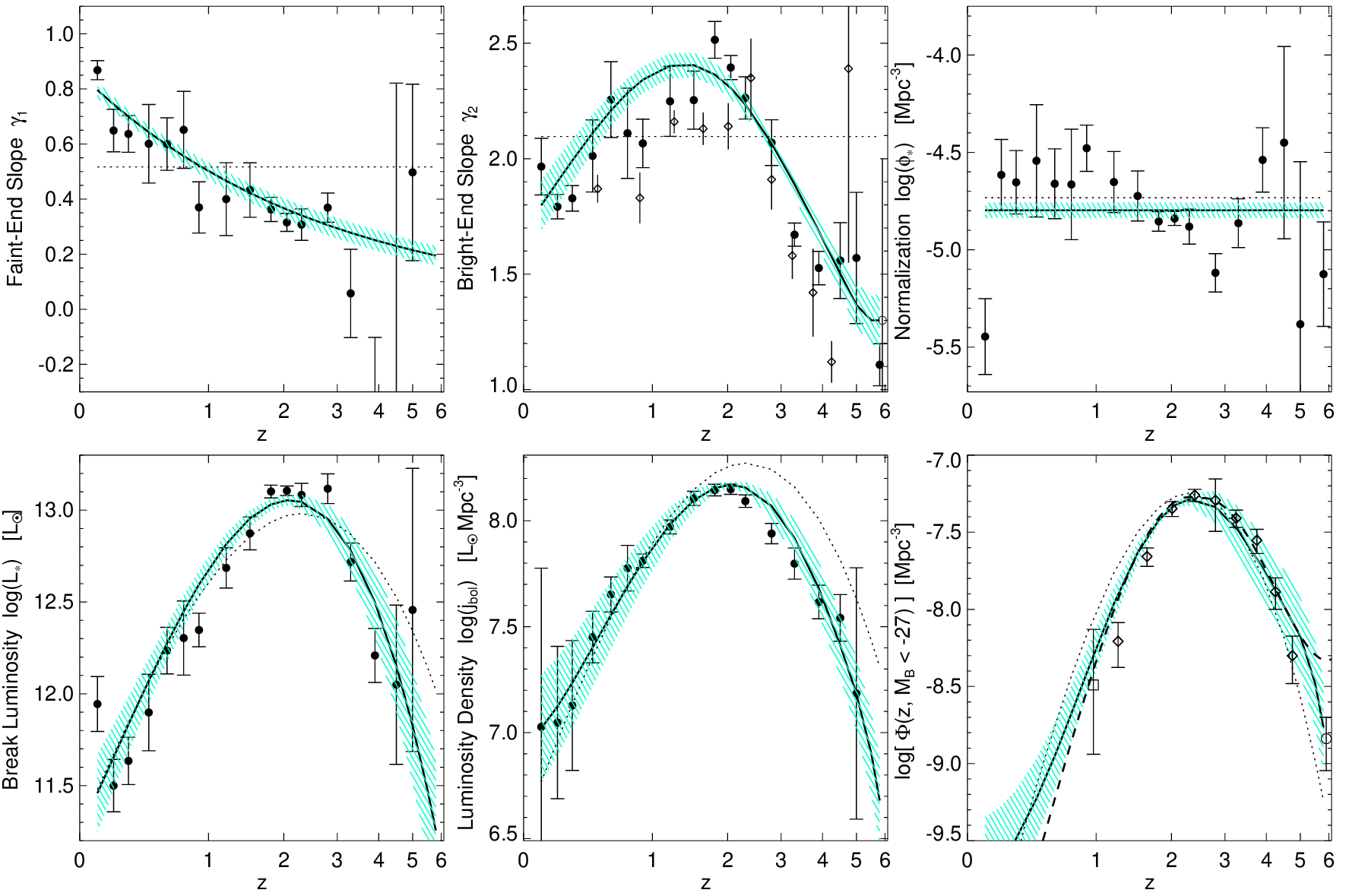}
    \caption{The best-fit QLF double-power law parameters as a function of redshift. 
    Points show the best-fit values to data at each redshift, dotted lines the best-fit 
    PLE model, 
    and solid lines the best-fit full model (with cyan shaded range showing the 
    $1\sigma$ uncertainty). Open diamonds in $\slopebright(z)$ show the bright-end 
    slope fits from \citet{DR3}. Although PLE is appropriate 
    for a lowest-order fit, both the bright and faint-end slopes evolve with redshift 
    to high significance ($>6\sigma$). Lower right shows the predicted number density 
    of bright optical ($M_{B}<-27$) quasars from the full fit (solid), compared to that 
    observed in \citet{Croom04} (square), \citet{DR3} (diamonds and 
    dashed line), and \citet{Fan04} (circle). 
    \label{fig:fitted.params}}
\end{figure*}

We plot the best-fit parameters from these fits at a number of
redshifts in Figure~\ref{fig:fitted.params} 
(error bars show formal $1\sigma$ uncertainties from the fits),
and list them at several redshifts in Table~\ref{tbl:fits.at.z}.  The
normalization $\phistar$ is roughly constant, while the break
luminosity (which is quite tightly constrained for most of the
redshift range) evolves by $\sim2$ orders of magnitude. Note that the
points where $\phistar$ appears to deviate from being constant also
have discrepant $\lstar$ values --- the degeneracy between the two is
such that the value of $\phistar$ is consistent with being constant at
all $z$ for smooth evolution in $\lstar$.  There is an indication of
evolution in both the faint-end and bright-end slopes, which we
discuss below.  The integrated bolometric luminosity density is well
constrained, with the largest uncertainties only $\sim0.15\,$dex. 
We determine the luminosity density by integrating the best-fit 
luminosity function to $L=0$. At most redshifts $z\gtrsim0.5$ the 
faint-end slope is relatively shallow, and choosing instead a cutoff at e.g.\ 
$L=10^{8}-10^{9}\,L_{\sun}$ changes the integrated 
luminosity density by $\lesssim10\%$. At the lowest redshifts, 
however, the faint-end slope is steep, and at $z=0$ where this is 
most pronounced, the luminosity 
density is $\sim15\%$ lower if we truncate at $L=10^{8}\,L_{\sun}$ 
($\sim25\%$ lower for $L=10^{9}\,L_{\sun}$). This sensitivity to 
the steep faint-end slope at low-$z$ is the reason for the relatively large 
uncertainty in the luminosity density at low redshift. At
high-$z$ the uncertainty owes to the limited amount of data. 
In any case there is
a well-defined peak in the luminosity density at $z= 2.154\pm 0.052$ 
(formal error from fit; we expect a 
systematic error $\pm0.15$ from choices in sampling and binning 
the observations), well 
outside the range where either of these systematic concerns is problematic.

\subsection{Analytic Fits as a Function of Redshift}
\label{sec:fits}

We characterize the QLF as a function of redshift by adopting
a standard pure luminosity evolution (PLE) model, where the bolometric
QLF is a double power law at all $z$, with constant $\slopefaint$,
$\slopebright$, and $\phistar$, but an evolving $\lstar$.  We allow
$\lstar$ to evolve as a cubic polynomial in redshift,
\begin{equation}
\log{\lstar} = (\log{\lstar})_{0} + k_{L,\,1}\,\xi + k_{L,\,2}\,\xi^{2} + k_{L,\,3}\,\xi^{3} \, ,
\end{equation}
where 
\begin{equation}
\xi = \log{\Bigl (}\frac{1+z}{1+z_{\rm ref}}{\Bigr)}.
\end{equation}
Here, $k_{L,\,1}$, $k_{L,\,2}$, and $k_{L,\,3}$ are free parameters,
and we set $z_{\rm ref}=2$ (which roughly minimizes the covariance in
the fit). The cubic term is demanded by the data
($\Delta\chi^{2}\sim 600$ on its addition), but higher order terms in
$\xi$ are not ($\Delta\chi^{2}\lesssim 1$). Since this model
includes the evolution with redshift, we can simultaneously fit to all
the data sets in Table~\ref{tbl:qlfs}, each over the appropriate
redshift intervals of the observed samples.

The best-fit PLE model parameters are given in Table~\ref{tbl:params}
and plotted as a function of redshift in
Figure~\ref{fig:fitted.params}, and the resulting QLF
is shown in Figures~\ref{fig:show.z.low} \& \ref{fig:show.z.high}. 
Although this provides a reasonable lowest-order
approximation to the data, it fails at the faint end, underpredicting
the abundance of low-luminosity sources at $z\lesssim0.3$ and
overpredicting it at $z\gtrsim 2$, and the fit is poor at $z\gtrsim 5$
with much too steep a bright end slope. Over the entire data set, the
fit is unacceptable, with $\chi^{2}\approx1924$ for $\nu=510$ degrees
of freedom.  A pure density evolution (PDE) model fares even worse,
with $\chi^{2}/\nu=3255/510$ (although for completeness we provide the
best-fit PDE parameters in Table~\ref{tbl:params.ldde}), unsurprising
given that nearly every observed data set which resolves the break in
the QLF favors the PLE form
\citep[e.g.,][]{Boyle00,Miyaji00,Ueda03,Croom04,2SLAQ}.

As discussed in \S~\ref{sec:intro}, many recent 
studies have found evidence for evolution in the faint
end beyond that predicted by PLE, with the density of lower-luminosity
sources peaking at lower redshift than the density of
higher-luminosity sources, and have fit this trend with a
luminosity-dependent density evolution (LDDE) model
\citep{SchmidtGreen83}, while high-redshift samples have suggested 
evolution in the bright-end QLF shape, confirmed robustly as a 
flattening in $\gamma_{2}$ for the 
first time in a homogeneous sample by the SDSS \citep{Richards06}.
We wish to incorporate this
additional evolution into our model.

For comparison with these results and future X-ray surveys, we fit to 
an LDDE form allowing maximal flexibility of the parameters, 
\begin{eqnarray}
\phi(L,\,z)&=&\phi(L,\,0)\,e_{\rm d}(L,\,z)\\
&=&\frac{\phi_{\ast}}{(L/L_{\ast})^{\gamma_{1}}+(L/L_{\ast})^{\gamma_{2}}}\,e_{\rm d}(L,\,z). 
\end{eqnarray}
The density function 
$e_{\rm d}$ is given by 
\begin{equation}
  e_{\rm d}(L,\,z) = \left\{ \begin{array}{ll}
      (1 + z)^{p1}  & (z\leq z_{c}) \\
      (1 + z_{c})^{p1}\,[(1+z)/(1+z_{c})]^{p2} & (z>z_{c})
\end{array}
    \right.
\label{eqn:LDDE.density.function}
\end{equation}
with 
\begin{equation}
  z_{c}(L) = \left\{ \begin{array}{ll}
      z_{c,\,0}(L/L_{c})^{\alpha}  & (L\leq L_{c}) \\
      z_{c,\,0} & (L>L_{c})
\end{array}
    \right.
\end{equation}
and 
\begin{eqnarray}
& & p1(L)=p1_{46}+\beta_{1}\,[\log(L/10^{46}\,{\rm erg\,s^{-1}})]\\
& & p2(L)=p2_{46}+\beta_{2}\,[\log(L/10^{46}\,{\rm erg\,s^{-1}})] .
\end{eqnarray}
Note that some authors \citep[e.g.,][]{HMS05} adopt an alternative normalization 
convention in terms of 
$A_{44}\equiv \phi(L=10^{44}\,{\rm erg\,s^{-1}},\, z=z_{c})$
and $z_{c,\,44}\equiv z_{c}(L=10^{44}\,{\rm erg\,s^{-1}})$, 
but these choices are not particularly convenient for 
the bolometric QLF. 
The 11 free parameters in this fit are then $\lstar$, $\phi_{\ast}$, $\slopefaint$, 
$\slopebright$, $z_{c,\,0}$, $L_{c}$, $\alpha$, $p1_{46}$, 
$p2_{46}$, $\beta_{1}$, $\beta_{2}$. Their best-fit values 
are given in Table~\ref{tbl:params.ldde}.

The LDDE form can effectively describe evolution in the faint-end QLF slope, 
but it does not allow for evolution in the bright-end slope. It also 
has a tendency to introduce a ``second break'' in the faint end of the QLF (i.e.\ if 
the faint end flattens there is often some $L$ below which it rises steeply again), 
for which we see no evidence. Ultimately, the improvement over the PLE fit is 
highly significant, with $\chi^{2}/\nu=1389/507$, but 
there is still room for substantial improvement. 

Therefore, we instead consider the PLE form above, but allow both the bright- and faint-end
slopes to evolve with redshift.  For the faint-end slope, using the fitted points at each $z$
to inform our choice of functional form, we model $\slopefaint$ as a
power-law in redshift,
\begin{eqnarray}
\slopefaint &= (\slopefaint)_{0}\times10^{k_{\slopefaint}\,\xi} \\
&= (\slopefaint)_{0}\,{\Bigl(}\frac{1+z}{1+z_{\rm ref}}{\Bigr)}^{k_{\slopefaint}} \, ,
\end{eqnarray}
where again $z_{\rm ref}=2$ is fixed.  Allowing this dependence (while
still holding $\slopebright$ constant) significantly improves the
quality of the fit relative to the PLE model,
reducing $\chi^{2}$ by $\Delta\chi^{2}\approx 500$
with the addition of one parameter, to $\chi^{2}/\nu=1422/510$. The
values for this fit are given in
Table~\ref{tbl:params}. There is no evidence for higher-order terms
($\Delta\chi^{2}\lesssim1$ for the addition of a second-order power in
$\xi$). We have also tested different functional forms, and do not
find any which provide a significantly better fit.  This choice also has
the advantage that it extrapolates to a flat $\slopefaint\rightarrow0$
(as opposed to a negative, likely unphysical) slope at high redshift.

Parameterizing evolution in the bright-end slope is more difficult. There
appears to be evidence for a steepening of the bright-end slope
(increase in $\slopebright$) from $z\sim0$ to $z\sim1.5$, then a
flattening with redshift.  We find the best results for a double-power
law of the form
\begin{eqnarray}
\slopebright &=& (\slopebright)_{0}\times 2\,{\Bigl[}
10^{k_{\slopebright,\,1}\,\xi} + 10^{k_{\slopebright,\,2}\,\xi}{\Bigr]}^{-1} \\
&=& \frac{2\,(\slopebright)_{0}}{{\Bigl(}\frac{1+z}{1+z_{\rm ref}}{\Bigr)}^{k_{\slopebright,\,1}}
+ {\Bigl(}\frac{1+z}{1+z_{\rm ref}}{\Bigr)}^{k_{\slopebright,\,2}}}.
\end{eqnarray}
Here, $k_{\slopebright,\,1}$ describes the rise of $\slopebright$ with
$z$ at low redshift, and $k_{\slopebright,\,2}$ describes the fall of
$\slopebright$ with $z$ at high redshift. The factor of $2$ is
inserted such that if $k_{\slopebright,\, 1}=k_{\slopebright,\, 2}=0$,
the resulting $\slopebright=(\slopebright)_{0}$.  The best-fit
parameters, fixing $\slopefaint$ or allowing both slopes to evolve
simultaneously, are given in Table~\ref{tbl:params}.  We find that
$k_{\slopebright,\,1}$ is non-zero at high formal significance, but
cannot say whether this steepening up to $z\sim1.5$ is ``real'' in any
robust sense; however, including this parameter does significantly
improve the accuracy of our fitting function in representing the
binned data.  The flattening at high redshifts is more significant
($\sim7\sigma$).  Including these two parameters greatly improves the
quality of the fit by $\Delta\chi^{2}\approx 600$, giving
$\chi^{2}/\nu = 1312/509$ for a fit with constant $\slopefaint$ or
$\chi^{2}/\nu = 1007/508$ for a simultaneous fit including evolution in
$\slopefaint$ and $\slopebright$.  Again, there is no evidence for
terms describing further evolution.
Although our final best fit does not cross
this limit until the highest redshifts, we also formally enforce a
lower limit $\slopebright\geq 1.3$ (to prevent an unphysical
divergence in the luminosity density).

The best-fit parameters as a function of redshift (allowing all free
parameters to vary) are plotted in
Figure~\ref{fig:fitted.params}. There is a slight offset between the $\lstar$ and
$\phistar$ of this fit and that fitted at specific redshifts, 
but this owes to their covariance. This can be seen in 
the bolometric luminosity density, which 
accurately traces the fit predictions (note that the
PLE model significantly overpredicts the luminosity density at $z>2$,
as it overpredicts the number of faint sources).  
Our restriction $\slopebright\geq1.3$ is important in extrapolating
beyond $z\sim5$ to prevent a divergence in the number and luminosity
density of bright sources. We also plot the predicted number density
of quasars with $M_{B}<-27$ as a function of redshift, which 
agrees with the direct observations (expected, but nevertheless a 
reassuring consistency check). 

\begin{figure*}
    \centering
    \plotone{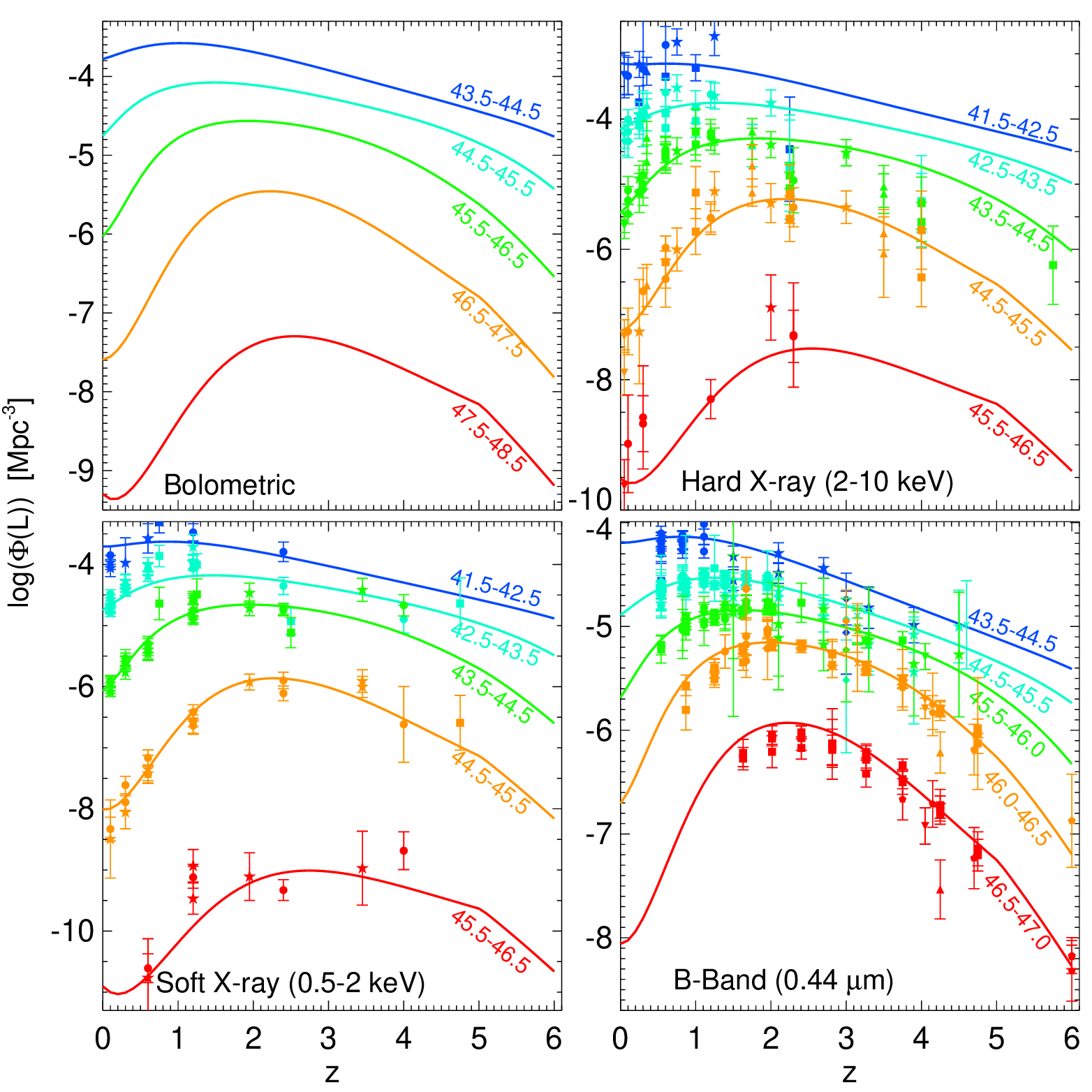}
    \caption{Total number density of quasars in various luminosity intervals 
    (in $\log{(L_{\rm band}/[{\rm erg\,s^{-1}}])}$ as labeled)
    as a function of redshift, from the best-fit evolving double-power law model 
    (lines) and the compiled observations in Table~\ref{tbl:qlfs} (points of    
    corresponding colors, symbols for each sample listed therein) 
    in bolometric luminosity, $B$-band, soft X-rays ($0.5-2$\,keV), and  
    hard X-rays ($2-10$\,keV). The trend that the density of lower-luminosity 
    AGN peaks at lower redshift is manifest in all bands. 
    \label{fig:n.l.z}}
\end{figure*}

Figure~\ref{fig:n.l.z} plots the number density of 
quasars integrated over various luminosity intervals, in various bands as a function of redshift, 
from the best-fit model and the compiled observations in Table~\ref{tbl:qlfs}. 
Although this information is contained 
in Figures~\ref{fig:show.z.low} and \ref{fig:show.z.high}, Figure~\ref{fig:n.l.z} 
nicely illustrates an essential trend captured by the LDDE or evolving double power-law forms, 
namely that the density of lower-luminosity sources 
peaks at lower redshift than that of higher-luminosity sources. The trend is evident 
in all bands we consider.
    
We briefly consider an alternate model for the shape of the QLF,
which illustrates several important points. Instead of the double
power-law of Equation~(\ref{eqn:doublepwrlaw}), we consider a modified
Schechter function
\begin{equation}
\phi(L)={\phistar}\,{(L/\lstar)^{-\slopefaint}}\,
{\exp{{\Bigl\{}-{\Bigl(}\frac{L}{\lstar}{\Bigr)}^{\slopebright}{\Bigr\}}}}.
\label{eqn:expshape}
\end{equation}
We have also, for example, adopted polynomials of
arbitrarily high order (although the fits typically do not improve
beyond fourth order). In either case, the fit is quite similar at 
most luminosities (with similar $\chi^{2}/\nu$), implying 
that there is no dramatic shape dependence which we have 
not captured. However, such a fit does exhibit smoother curvature 
rather than a sharp break at $L_{\ast}$, evidence for which has been seen 
in some optical samples \citep[e.g.,][]{COMBO17,2SLAQ}. 
At the highest luminosities ($\gtrsim1$\,dex above $\lstar$, typically
$L_{\rm bol}\gtrsim10^{14}\,L_{\sun}$) the implied number of quasars
is an order of magnitude lower for these parameterizations than for the double power law
prediction, and falls much more rapidly. 
The resulting observed luminosity functions are more sensitive to the estimated dispersion in
bolometric corrections, but in either case the highest luminosity
soft and hard X-ray objects in Figures~\ref{fig:show.z.low} \&
\ref{fig:show.z.high} are substantially affected by quasars shifting
into slightly larger bins of soft or hard X-ray luminosity owing to
different spectral shapes (e.g.\ the scatter in $\alpha_{\rm ox}$). It
is important to account for this effect when attempting to infer the
number density of the most massive black holes and most luminous
quasars, as a naive extrapolation of the median bolometric corrections
applied to the most X-ray bright quasars implies extreme (and
potentially unphysical) bolometric luminosities
$\gtrsim10^{15}\,L_{\sun}$ (i.e.\ a $\gtrsim3\times10^{10}\,M_{\sun}$
black hole at the Eddington rate). Multiwavelength observations of
these particular objects and further study from large area
surveys which do not have to bin in widely spaced luminosity intervals
will be critical in breaking the degeneracies between these fits to the
intrinsic bolometric QLF and the double power-law form.

We are unable to find any further dependences which significantly
improve the best-fit QLF.  Allowing the parameters describing the
observed column density distribution and spectral shape, and their
respective luminosity dependence, to vary simultaneously, yields only
marginal improvement. It
appears that the remaining scatter in the data does not mostly owe to
a failure to capture some remaining dependence. In
Table~\ref{tbl:qlfs} we list the $\chi^{2}/\nu$ for each sample with
respect to the best-fit full model of the evolution of the QLF in
Table~\ref{tbl:params}. The agreement with most samples is good, and
the largest, most well-constrained samples (with small typical $\ll
0.1\,$dex errorbars) give $\chi^{2}/\nu\lesssim3$, in each case
comparable to or smaller than the reduced $\chi^{2}$ found by the
respective authors in fits to those {\em individual} data sets. To the
extent that the functional forms adopted by these authors cannot
reduce the variance below this level, we would not expect to do better
in the combined sample.

There will also be some unavoidable variance introduced owing to
systematic variations between independent data sets.  Various
observational calibrations, the model dependences inherent in
calculating a binned QLF, and most of all cosmic variance will all
contribute to sample-to-sample differences.  In fact, we find that
allowing for even a small $\sim0.05$\,dex ($10-15\%$) systematic
normalization variance between samples, 
most of the remaining scatter is accounted for, with a
best-fit model improvement $\Delta\,\chi^{2}\sim 500$.  We provide the
values of this fit in Table~\ref{tbl:params}, but caution that we have
increased the sample-to-sample variance by this amount uniformly, when
in fact systematic effects such as cosmic variance will be smaller for
large surveys such as the SDSS and 2dF than for the small deep fields
of Chandra and XMM.  Consequently, this under-weights some of the most
well-constrained observations, and the fit results should be
considered to be heuristic.

\section{The Buildup of the Black Hole Population}
\label{sec:lifetime}

\subsection{Model-Dependent Quantities}
\label{sec:models}

We can gain further insight into the evolution of the QLF, albeit at
the cost of some model dependence, by de-convolving the observed
quasar luminosity function with a theoretical model for the quasar
light curve or lifetime.  This method is well-established
\citep[e.g.,][]{Salucci99,YT02,Marconi04,Shankar04,YL04}, but most
studies adopt simplified ``toy'' light curves with accretion at a
constant (fitted) absolute rate or Eddington ratio and fitted
lifetimes and duty cycles.  We instead follow \citet{H06b}, who derive
physically motivated quasar light curves from simulations of merging
galaxies that include black hole growth (e.g. Springel et al. 2005b)
and which are consistent with a
large range of observational constraints that cannot be reproduced by
idealized models. This also removes the various fitting degeneracies
-- for a given bolometric QLF, the quasar light curves determined in
simulations yield a unique black hole mass function, cosmic X-ray
background spectrum, and self-consistent black hole and host galaxy
properties.

Given a consistent model of the quasar lifetime/lightcurve, the
observed bolometric QLF is given by the convolution of the rate of
quasar formation or ``triggering'' with the differential quasar
lifetime,
\begin{equation}
\phi(L) = \int{\dot{\phi}(M_{BH})\,\frac{{\rm d}t}{{\rm d}\log{L}}(L\,|\,M_{BH})\,{\rm d}\log{M_{BH}}}
\end{equation}
where 
\begin{equation}
\dot{\phi}(M_{BH}) = \frac{{\rm d}\Phi (M_{BH},\,t)}{{\rm d}\log{M_{BH}}\,{\rm d}t}
\end{equation}
is the rate of formation of black holes of a relic mass $M_{BH}$ at
cosmic time $t$. Since ${\rm d}t/{\rm d}\log{L}$ is completely
determined in the simulations and analytical models of
\citet{H06a,H06b}, we can fit to $\dot{\phi}(M_{BH})$ in the same
manner that we have fitted $\phi(L)$.

We assume a double power-law form 
for $\dot{\phi}(M_{\rm BH})$ at all $z$, 
\begin{equation}
\dot{\phi}(M_{\rm BH}) = \frac{\dot{\phi}_{\ast}}
{(M_{\rm BH}/M_{\ast})^{\eta_{1}}+(M_{\rm BH}/M_{\ast})^{\eta_{2}}}
\end{equation}
with normalization $\dot{\phi}_{\ast}$, break $M_{\ast}$, 
and faint-end and bright-end slopes $\eta_{1}$ and $\eta_{2}$, respectively. 
We determine an analytical fit to $\dot{\phi}(M_{\rm BH},\, z)$
in the same manner as $\phi(L,\,z)$. We allow 
$\log{M_{\ast}}$ to vary as a cubic in $\xi$ just as $L_{\ast}$, 
and allow the high-mass slope $\eta_{2}$ to vary just as $\slopebright$, as well. 
There is no evidence for evolution of the low-mass slope $\eta_{1}$. 
In this formulation there is evidence for evolution in
$\dot{\phi}_{\ast}$ above $z\sim2$, so we fix it to a constant 
below $z_{\rm ref}\equiv2$, and allow it to evolve as a power-law above $z_{\rm ref}$, 
\begin{equation}
  \dot{\phi}_{\ast}(z) = \left\{ \begin{array}{ll}
      (\dot{\phi}_{\ast})_{0}  & (z\leq z_{\rm ref}) \\
      (\dot{\phi}_{\ast})_{0}\,[(1+z)/(1+z_{\rm ref})]^{k_{\dot{\phi}}} & (z> z_{\rm ref}).
\end{array}
    \right. 
\end{equation}

\begin{figure*}
    \centering
    \plotone{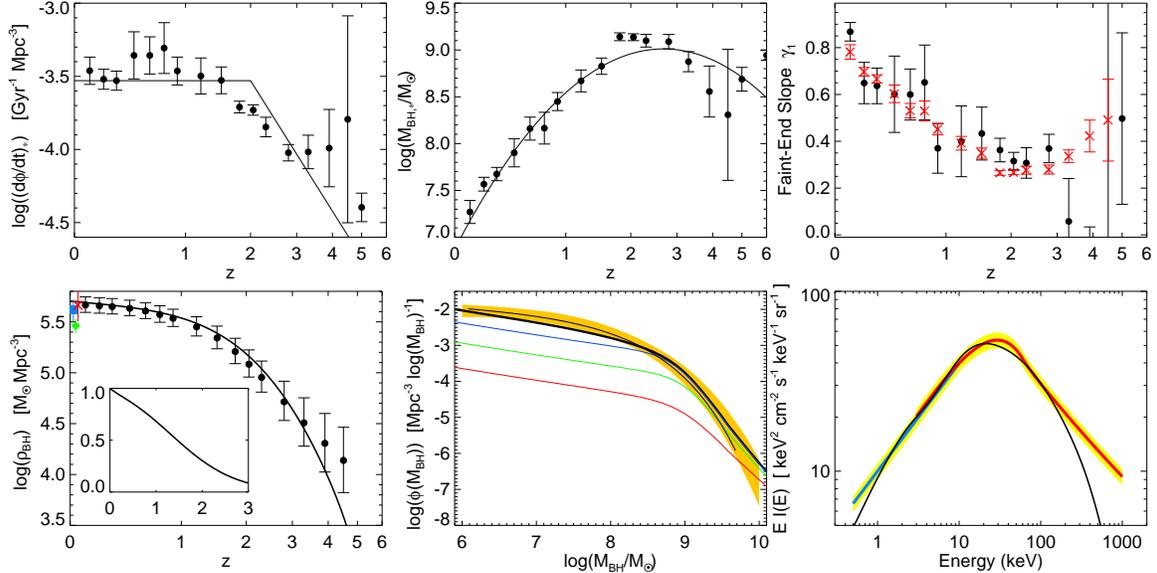}
    \caption{Best-fit parameters to $\dot{\phi}(M_{\rm BH})$, the ``birthrate'' or 
    formation rate of black holes at a given redshift, given the 
    quasar lifetimes determined in the simulations of \citet{H06a,H06b}. 
    Top left shows $d\phi/d{\rm t}$, the quasar activation / birthrate 
    normalization. Top center shows $M_{\ast}$, the break in 
    the characteristic final black hole mass forming at a given $z$. 
    Points are fitted at each redshift, 
    solid lines show best-fit from $\dot{\phi}(M_{\rm BH})$ as a function of redshift. 
    Top right shows the 
    faint-end slope of the observed QLF from Figure~\ref{fig:fitted.params} 
    at each redshift (black circles), 
    with the predicted faint-end slope from the same quasar lifetime models 
    (red crosses determined from the observations at each $z$).
    Bottom left shows the total black hole mass density of the Universe, 
    from the data (circles), best fit QLF (solid line). Inset shows 
    the fraction of the $z=0$ mass density ($\rho_{\rm BH}(z)/\rho_{\rm BH}(0)$), 
    on a linear scale. Green diamond, blue square, and red cross show the 
    local black hole mass density calculated in \citet{YT02,Marconi04}, 
    and \citet{Shankar04}, respectively. 
    Bottom center shows the integrated black hole mass function 
    at $z=0$ (thick black line) and $z=1,\ 2,\ 3$ (blue, green, and red 
    lines, respectively), compared to the inferred local mass function from 
    \citet{Marconi04} (yellow shaded range shows $1\sigma$ range) and 
    \citet{Shankar04} (thin black line). Bottom right shows the integrated 
    cosmic X-ray background (black line), compared to the observations of 
    \citet{Barcons00} and \citet{Gruber99} (blue and red lines, respectively, 
    with shaded yellow $1\sigma$ observational uncertainty). 
    \label{fig:fits.model}}
\end{figure*}

In Figure~\ref{fig:fits.model}, we plot the best-fit normalization
$\dot{\phi}_{\ast}$ and break (characteristic mass in formation)
$M_{\ast}$ as a function of redshift. At
high redshifts, objects build up rapidly, until $z\sim2$, after
which the rate of merger/black hole growth ``events'' flattens, and
activity ceases in the most massive systems and rapidly moves to
less massive objects at lower redshifts, perhaps driven by feedback
mechanisms quenching activity in the higher-mass systems.  Given the
adopted lifetime models, the observed faint-end $\phi(L)$ slope
$\slopefaint$ is dominated by sources with $L\ll L_{\rm Edd}(M_{\rm
BH})$ and is determined by the quasar lifetime as a function of
$M_{\rm BH}$. Figure~\ref{fig:fits.model} shows the
predicted $\slopefaint(z)$ from the quasar lifetime model given the
best-fit $\dot{\phi}(M_{\rm BH})$ at each $z$, compared
to the direct fits from Figure~\ref{fig:fitted.params};
the agreement is good despite $\eta_{1}$ being nearly constant and
much flatter $\eta_{1}\approx 0.0-0.2$ at all redshifts (see also
Figure~3 of \citet{H06a}).

\subsection{Integrated Quantities}
\label{sec:integrate}

If all black holes accrete with some constant radiative 
efficiency $\epsilon_{r}$, 
\begin{equation}
L_{\rm bol} = \epsilon_{r}\,\dot{M}\,c^{2}
\end{equation}
then the integrated black hole mass density is given by integrating 
the luminosity density \citep{Soltan82}. 
Figure~\ref{fig:fits.model} shows the result of this integration from $z\rightarrow\infty$ to $z$, 
assuming $\epsilon_{r}=0.1$.
We extrapolate beyond $z=6$ but this only changes the $z=0$ result by $\approx0.03$\,dex.
Note that the vertical axis is logarithmic; 
on a linear scale most of the growth occurs at $z\lesssim1-2$, in agreement 
with previous estimates \citep[e.g.,][]{YT02,Marconi04,Shankar04}. 

This gives a $z=0$ black hole mass density of 
\begin{equation}
\rho_{\rm BH}(z=0) = {4.81}^{+1.24}_{-0.99}\,{\Bigl(}\frac{0.1}{\epsilon_{r}}{\Bigr)}\,h_{70}^{2}\times 10^{5}\,M_{\sun}\,{\rm Mpc^{-3}}.
\end{equation}
Comparing this to calculations from local bulge mass, luminosity, and velocity dispersion 
functions: $\rho_{\rm BH}=4.2\pm1.1$ \citep{Shankar04}, 
$\rho_{\rm BH}=4.6^{+1.9}_{-1.4}$ \citep{Marconi04}, $\rho_{\rm BH}=2.9\pm0.5$ 
\citep{YT02} (each in units of $h_{70}^{2}\times 10^{5}\,M_{\sun}\,{\rm Mpc^{-1}}$), 
the agreement is good. 
The dominant sources of uncertainty are systematic: Tables~\ref{tbl:params} and 
\ref{tbl:params.ldde} show the integrated $\rho_{\rm BH}(z=0)$, 
demonstrating that different fitting functions with similar 
$\chi^{2}/\nu$ (or e.g., allowing 
$L_{\ast}$ to evolve as a polynomial in $z$ instead of $\xi$) yields similar 
factors $\sim1.5-2$ systematic differences in $\rho_{\rm BH}(z=0)$. 
An additional contribution from Compton-thick sources is another 
uncertainty, although it cannot be much larger or would overpredict the 
present mass density and 
X-ray background (below). Fractional uncertainties in $\epsilon_{r}$ are 
substantial, and e.g.\ \citet{HNH06} 
estimate based on radiatively inefficient accretion flow models 
\citep[e.g.,][]{NY95} a mass-weighted 
effective $\epsilon_{r}$ a factor of $\sim0.8$ lower than the ``radiatively 
efficient'' value owing to some growth at low Eddington ratio. 
Given these uncertainties, it is not necessarily meaningful to 
compare with the local black hole mass density with any greater statistical accuracy. 

Having factored out the quasar lifetime in \S~\ref{sec:models} to determine 
the rate of build-up of {\em individual} black holes $\dot{\phi}(M_{\rm BH})$, 
we can trivially integrate this over redshift to 
determine the relic black hole mass function 
\citep[see][for a full derivation of the relevant equations]{H06b}. We show 
the resulting $z=0$ black hole mass function in Figure~\ref{fig:fits.model}, 
compared to the local mass function estimates from 
\citet{Shankar04} and \citet{Marconi04}, determined from 
local bulge velocity dispersion, mass, and luminosity 
function observations  
\citep{Marzke94,Kochanek01,Cole01,Nakamura03,Blanton03,Bernardi03,Sheth03}. 
The integrated black hole 
mass function agrees well at all masses. We emphasize that there are no 
free parameters to fit to the black hole mass function in this analysis, 
as there are in the traditional analyses of e.g.\ \citet{YT02}. 
The mass functions at several redshifts are also shown: the 
highest-mass black holes are formed at redshifts $z\sim2$, and low mass black holes 
assemble at $z\lesssim1$, in agreement with a number of previous studies 
\citep[e.g.,][]{Merloni04,Marconi04,Shankar04,H06b}. 

Finally, we can integrate over the bolometric QLF, including our full modeling of 
spectral shapes and column density distributions 
to determine the X-ray background spectrum
\citep[see also][]{Ueda03,H06b}. Figure~\ref{fig:fits.model} 
compares this with the observed X-ray background spectrum from 
\citet{Barcons00} at $E<10\,{\rm keV}$ and \citet{Gruber99} at $E>3\,{\rm keV}$ 
(normalized to the same amplitude 
in the range of overlap based on the detailed compilation in Barcons et al.\ 2000) 
with the final $\sim10\%$ absolute normalization uncertainty shown as the yellow range. 
The predicted X-ray background spectrum agrees well with that observed over 
the entire $\sim1-100$\,keV range over which ``normal'' AGN spectra are expected to 
dominate the X-ray background contribution, and combined 
with the black hole mass density constraint implies that the X-ray background 
is dominated by AGN with canonical radiative efficiencies 
$\epsilon_{r}\sim0.1$ \citep[compare e.g.][]{ERZ02}. This also puts strong 
limits on additional contributions of Compton-thick quasars, as a fraction much larger 
than the (luminosity-dependent) $\sim30\%$ of \citet{Ueda03} would be problematic.

\subsection{UV and Ionizing Backgrounds}
\label{sec:UV}

\begin{figure*}
    \centering
    \plotone{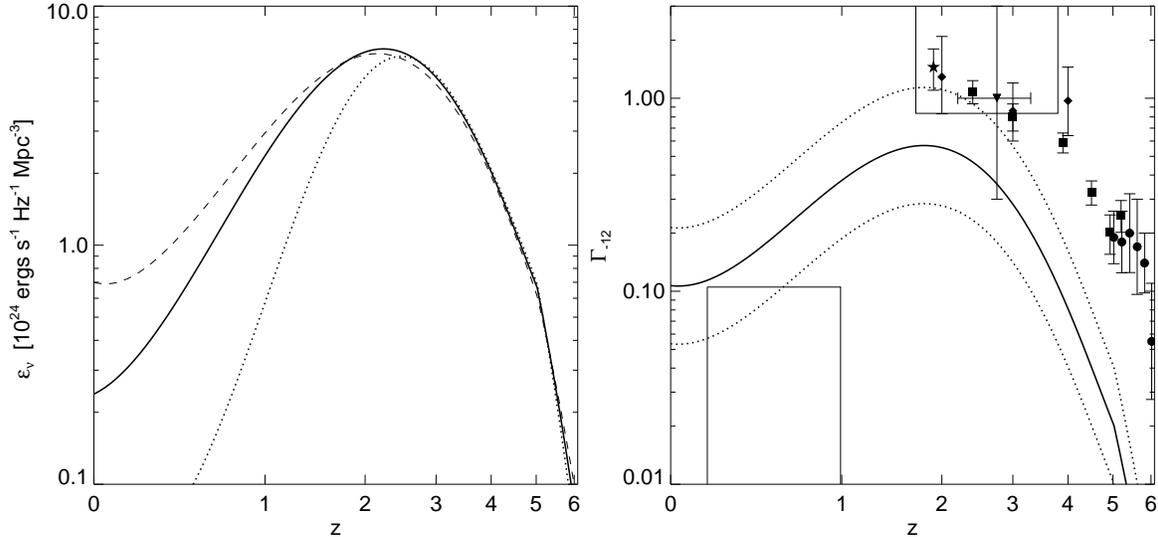}
    \caption{Left: Specific luminosity density at $912\,$\AA, calculated using our full 
    spectral and obscuration modeling (solid line). If instead we adopt a constant 
    conversion from the bolometric luminosity density in Figure~\ref{fig:fitted.params} 
    (or, similarly, ignore obscuration) the dashed line is obtained. 
    Taking a constant 
    conversion from the number density of quasars with $M_{B}<-27$ 
    (Equation~\ref{eqn:eps.proxy}; 
    here normalized to give the exact specific luminosity density at $z=3$) 
    yields the dotted line. Right: Implied rate of ionization $\Gamma_{-12}$ 
    (in units of $10^{-12}\,s^{-1}$) as a function of 
    redshift, from our full modeling (black line), adopting the conversions 
    from $\epsilon_{\nu}$ from \citet{Fardal98,SchirberBullock03}. Dotted lines show
    the same, but arbitrarily increased (decreased)
    by a factor of 2. Observational estimates of $\Gamma_{-12}$ 
    are shown, from \citet{Bolton05} (diamonds at $z=2-4$), \citet{Tytler04} 
    (star at $z=1.9$), \citet{Rollinde05} (triangle at $z=2.75$), 
    \citet{MME01,MME03} (squares at $z=2.4-5.2$), \citet{Fan05}
    (circles at $z=5-6$), and \citet{Scott00} (boxes at $z\sim0-1$ and 
    $z\sim2-4$). Below $z\sim2$, the faint quasar contribution is important, 
    with the effects of obscuration and a changing faint end slope requiring 
    a detailed calculation of the specific ionizing flux. However, above $z\sim2$, 
    the flatter bright and faint end slopes imply that bright quasars dominate the 
    luminosity density, and the evolution in the number density of bright quasars becomes an 
    accurate proxy for the evolution in $\epsilon_{\nu}$. Quasars can account 
    for all of the observationally estimated $\Gamma_{-12}$ at $z\lesssim1$, 
    and $\sim50\%$ at $z\sim2-3$ (for the specific conversions adopted), but 
    the quasar luminosity density drops off much faster than $\Gamma_{-12}$ at higher $z$.
    \label{fig:uv}}
\end{figure*}

Figure~\ref{fig:uv} shows the specific luminosity density $\epsilon_{\nu}$ 
at $912\,$\AA, from our full spectral and obscuration modeling, 
integrated over our full best-fit QLF model at each redshift. Essentially, this is identical 
to integrating the $B$-band luminosity functions and using the 
conversion $L_{\nu}(912\,$\AA$)=10^{18.15}\,{\rm ergs\,s^{-1}\,Hz^{-1}}\,(L_{B}/L_{\sun})$, 
\citep{Elvis94,Telfer02,Richards06}. 
We also show the bolometric luminosity density from Figure~\ref{fig:fitted.params}, 
renormalized by a constant ``bolometric correction,'' and the $\epsilon_{\nu}$
implied if this luminosity density were proportional 
to the number density of bright quasars, $\epsilon_{\nu}\propto \Phi(M_{B}<-27)$, 
both normalized to give the inferred $\epsilon_{\nu}$ at $z=3$. The latter 
correction is given by 
\begin{equation}
\epsilon_{\nu}\approx0.95\times10^{24}\,{\rm ergs\,s^{-1}\,Hz^{-1}}\,
{\Bigl[}\frac{\Phi(M_{B}<-27)}{10^{-8}\,{\rm Mpc^{-3}}}{\Bigr]}.
\label{eqn:eps.proxy}
\end{equation}

At $z\lesssim2$, the faint and bright end slopes of the QLF are 
relatively steep, indicating an important contribution to luminosity 
densities from faint objects. As a result, $j_{\rm bol}$ (neglecting 
obscuration) and $\Phi(M_{B}<-27)$ (neglecting 
all faint objects) are poor proxies for $\epsilon_{\nu}$. Above 
$z\sim2$, however, the luminosity density is dominated by bright (and 
relatively unobscured) objects, and Equation~(\ref{eqn:eps.proxy}) provides an accurate 
approximation to the full calculation of the specific luminosity density from all available 
constraints in Table~\ref{tbl:qlfs}. 
 
Given $\epsilon_{\nu}$ at $912\,$\AA\ and some description of the 
IGM opacity, we can estimate the ionization rate contributed by the quasar 
population. The right panel of Figure~\ref{fig:uv} shows 
this rate $\Gamma_{-12}$ (rate in $10^{-12}\,{\rm ionizations\,s^{-1}\,atom^{-1}}$), 
where we have used the conversions from 
\citet{Fardal98,SchirberBullock03}
\begin{equation}
\Gamma_{-12}(z)\approx 2.0\,(1+z)^{-1.5}\,\frac{\epsilon_{24}}{3+|\alpha_{\rm UV}|}
\end{equation}
with $\alpha_{\rm UV}$ defined in \S~\ref{sec:bol.corr} and 
$\epsilon_{24}=\epsilon_{\nu}/(10^{24}\,{\rm ergs\,s^{-1}\,Hz^{-1}})$.
We compare with several observational estimates of the total $\Gamma_{-12}$ from 
\citet{Scott00,MME01,MME03,Tytler04,Bolton05,Rollinde05,Fan05} 
using various methods 
\citep[for a review, see][]{Fan06}. Note that this assumes a specific 
IGM optical depth, escape fraction, and Lyman Limit System distribution/evolution, which are by 
no means well-constrained, so we caution that we do not mean this to 
be a rigorous calculation of $\Gamma_{-12}$ and also show the same estimate multiplied (divided)
by a factor of $2$.
Ultimately, this rough calculation supports previous estimates 
\citep[e.g.,][]{MHR99,Fan01b,Bolton05} that at 
$z\sim2$, the ionizing background contributed by quasars is comparable to ($\sim50\%$)
that observed, but the quasar contribution 
drops much more rapidly with redshift than the ionizing background. The contributed 
$\epsilon_{\nu}$ declines from $z\sim2-6$ roughly $\propto(1+z)^{-4}$ (giving a
factor $\sim60$ drop in $\Gamma_{-12}$), 
whereas the observed total $\Gamma_{-12}$ falls by only a factor 
$\sim10$.

\section{Discussion/Conclusions}
\label{sec:discuss}

We have used a large set of observed quasar luminosity functions in
various wavebands (Table~\ref{tbl:qlfs}), from the IR through
optical, soft and hard X-rays, and emission line measurements,
combined with recent estimates of the quasar column density
distribution from hard X-ray and IR observations, and a large number of observations
from the radio through hard X-rays determining the distribution of quasar spectral
shapes to estimate the bolometric quasar luminosity function. This
allows us to fit the bolometric luminosity function over a wide
baseline in both luminosity and redshift, from bolometric luminosities
$\sim10^{41}-10^{49}\,{\rm erg\,s^{-1}}$ and redshifts $z=0-6$,
reliably determining the location of the break even at $z\sim3$ and
constraining its location with $\gtrsim2\sigma$ confidence up to
$z\sim4.5$.

Combining observations at different wavelengths but at the same
redshift allows us to test the self-consistency of measurements of the
column density distribution and spectral shape.  {\em With the
best-fit current estimates of the column density distribution and
quasar spectral shape, both of which depend on luminosity (the
unobscured fraction and optical/UV prominence in the quasar spectrum
increasing with luminosity), a single bolometric QLF
self-consistently reproduces the observed QLFs in each available band
at all redshifts for which we have compiled measurements}. A constant
(luminosity-independent) column density distribution cannot
self-consistently reproduce the observed QLF in different bands, 
and is increasingly ruled out by direct observations
\citep{HGD96,SRL99,Willott00,SR00,
Steffen03,Ueda03,GRW04,Hasinger04,SazRev04,Barger05,Simpson05,Hao05,
Sazonov06,Shinozaki06,Beckmann06b,Bassani06}.
A strong redshift dependence in the column density distribution is
also ruled out: measurements of such a dependence
\citep[e.g.,][]{LaFranca05} fit the observations with comparable
accuracy to a redshift-independent model up to $z\sim1$ (where the
observations are generally calibrated), but extrapolate poorly to
higher redshifts. 
Likewise, a single quasar spectrum cannot
self-consistently reproduce the observed QLFs in different
wavebands; it is necessary to account for 
the dependence of spectral shape on luminosity 
seen in \citet{Wilkes94,Green95,VBS03,Strateva05,Richards06,Steffen06}.

Extrapolating any fitted luminosity function outside of its
measured luminosity and redshift range can be inaccurate
by orders of magnitude (see, e.g.\ Figure~19 of \citet{DR3}), and we have 
demonstrated the importance of accounting for the detailed luminosity 
dependence of quasar SEDs and obscuration. It is also important to 
include the scatter in bolometric corrections, i.e.\ 
the fact that there is not one spectral shape even at a given
luminosity, as failure to account for this can
over-predict the number density of the most luminous quasars by an
order of magnitude, and over-predict the bolometric luminosity density
by $\sim30\%$. Given the large baselines spanned by our compiled  
samples and our full treatment of these effects, we provide a number of 
fitting formulae to simplify future comparisons, including: the bolometric 
QLF itself (Table~\ref{tbl:params}), median bolometric corrections 
(Equation~\ref{eqn:bolcorr}), 
the dispersion in bolometric corrections 
(Equation~\ref{eqn:bolcorr.scatter}), and effective 
``obscured/visible'' fractions (Equation~\ref{eqn:obsc}) as a 
function of luminosity. 

We find that the faint-end slope of the QLF flattens at increasing
redshift, at high significance ($>7\sigma$). Evidence for such a trend
comes from a number of different measurements
\citep{Page97,Miyaji00,Miyaji01,LaFranca02,Cowie03,
Ueda03,Fiore03,Hunt04,Cirasuolo05,HMS05}
(although we still find $\sim3\sigma$ evidence for this evolution
without these data), but independently these
samples detect such evolution with only marginal ($\lessim2\sigma$)
significance. The most significant
detection from a single study thus far has been that of \citet{HMS05},
who find an improvement from a $\sim5\%$ 2D K-S probability to
$\sim36\%$ allowing for faint-end evolution; it is simply difficult to 
obtain the large baseline necessary for strong constraints from a 
single sample. There has also 
been some debate over whether the steep faint-end slopes seen 
at $z\lesssim0.2$ by e.g.\ \citet{SazRev04,Hao05} are consistent with 
those seen in other (particularly optical) surveys, and indeed 
Figure~\ref{fig:show.z.low} demonstrates they are not consistent with the 
PLE model. However, accounting for the evolution in the 
faint end slope with redshift, Figure~\ref{fig:show.z.low} demonstrates that these observations
are consistent with the others we compile. 
Our analysis also allows us to understand the different
shapes of the QLF in different observed bands, and
Figures~\ref{fig:show.demo}, \ref{fig:show.z.low}, \&
\ref{fig:show.z.high} demonstrate that with obscuration more important
in low-luminosity quasars, the faint-end optical QLF is relatively
flattened at all redshifts. Together with optical samples generally
not probing to luminosities as faint as the X-ray samples, this
explains why the trend of evolution in the faint-end slope has been
preferentially observed in X-ray samples.

This confirms, at high significance, that there is faint-end evolution
{\em in the data}. This does not, of course, imply that the evolution
is necessarily real, because selection effects such as incompleteness
at low luminosities and high redshifts, or an evolving Compton-thick
population, could play a role. However, constraints on this quantity
are interesting. For example, to the extent that the evolution is
real, it supports at much greater significance the findings of
\citet{H06a}, namely the predicted functional dependence of the
faint-end shape on $\lstar$ (and, as a consequence, on redshift). This
model is motivated by hydrodynamic simulations 
(Hopkins et al. 2005a,b,c) and analytical models
of accretion rate evolution in feedback-driven outflows
(Hopkins \& Hernquist 2006),
which predict both that the faint-end shape should be controlled by
the shape of the quasar light curve, itself a function of quasar and
host galaxy properties.  To the extent that the faint-end slope does
or does not evolve, it directly constrains the quasar ``lifetime'' as
a function of luminosity \citep[i.e.\ differential time quasars spend
per luminosity interval in both growth and decay;][]{H05c}, and
consequently feedback coupling modes and their impact on the host
galaxy.

We also find at high significance ($>6\sigma$) that the {\em bright} end slope of
the QLF evolves with redshift, becoming shallower at high redshifts
$z\gtrsim3$.  This agrees with the early suggestions of
\citet{SSG95,Fan01a,Fan01b}, and subsequent more robust measurements of
\citet{DR3}.  We infer significant ($>3\sigma$) evidence for such evolution even
with those samples removed ($k_{\slopefaint,\,1}=1.22\pm0.15$,
$k_{\slopefaint,\,2}=-0.80\pm0.09$), but in this case the result is 
heavily influenced by the X-ray data, which at the highest
luminosities has a large bolometric correction ($L_{\rm bol}\sim
100\,L_{HX}$, from the observed $\alpha_{\rm ox}-L$ relation). Since
the QLF is falling steeply at these luminosities, a small change in
the bolometric correction will significantly change the inferred
slope; Figure~\ref{fig:show.demo} aptly demonstrates that the optical and IR
samples provide a much more robust probe of the ``true'' bright end
slope. Note that the best-fit trend at $z\sim6$ favors a shallower slope than 
recently suggested by \citet{Fan04,Cool06}, although the small 
number of objects make this only a $\sim1\sigma$ effect. Further 
wide-area surveys are needed to confirm the trend. 

There is also marginal evidence that the bright-end QLF slope
is somewhat shallow at $z\sim0$, steepening to $z\sim0.5-1$, then
becoming shallower with $z$ as described above.  This low-redshift trend is
marginally evident in the \citet{DR3} sample as well (see
Figure~\ref{fig:fitted.params}), although with a smaller low-$z$
baseline the authors assumed a constant bright-end slope up to
$z\sim3$.  There are a
number of potential systematic effects which complicate understanding
this trend.  For example, imperfect accounting of the (potentially
quite rapid) QLF evolution over a narrow redshift interval will bias
the luminosity function to a higher number density at large $L$, as will
many general binning procedures used in presenting the data, as well
as host galaxy contamination (see e.g.\ Figure~11 of Hao et al.\
2005). These problems are well known from local galaxy luminosity
functions, and can introduce significant effects even in samples
spanning the range e.g.\ $z\sim0-0.1$ \citep[see e.g.,][]{Blanton03},
where many of the low-redshift QLF measurements cover a significantly
larger $z\sim0-0.3$. Furthermore, the faint-end slope is observed to
be steep at $z\sim0$ \citep[e.g.,][]{Hao05}, implying a weak
break and large covariance between the best-fit parameters (much
larger than at any other redshifts where the break is constrained).
More careful re-analysis of these samples and e.g.\ recent 
local luminosity functions from \citet{Shinozaki06}, 
\citet{Sazonov06}, and \citet{Beckmann06b}, 
accounting for these effects, is needed.

Again, despite systematic uncertainties, 
evolution in the bright-end shape is interesting.  Models
predict that it should evolve with redshift, as a
consequence of feedback from AGN
\citep[e.g.,][]{KH00,WyitheLoeb03,Scannapieco05}.
At high redshift, with small galaxies,
large gas content, and rapid merger rates, the quasar activity is
expected to trace dark matter halo growth and merger rates, until a
redshift $z\sim2-3$ when feedback from quasar growth or other
mechanisms begins to dominate.
This ``shuts down'' activity, terminates
subsequent black hole growth and local star formation
\citep{H06c,SDH05a}, injects entropy \citep{Scannapieco04} and
radio-mode feedback from low accretion rate ``dying'' black holes
\citep{Croton06} heats the surrounding IGM and prevents new
infall, suppressing further galaxy and BH growth. In these models, this occurs first in
the largest systems, which evolve the most rapidly and have the
most massive, violent black holes, leaving a gas-rich merger and
corresponding quasar history which shifts to lower masses at lower
redshifts \citep{H06d}, and building up lower-mass elliptical galaxies
at preferentially later times \citep{H06c,H06d,H06e}, explaining observed
trends of ``cosmic downsizing'' \citep[e.g.,][]{Cowie96} in the
context of hierarchical structure formation.

Our estimate of the bolometric QLF allows us to immediately derive
several interesting quantities. The bolometric quasar luminosity
density (and correspondingly, total black hole accretion rate) of the
Universe follows a similar qualitative trend to that of $\lstar$,
peaking at a well-defined $z\approx2.154\pm0.052$. The integrated black
hole mass density to $z=0$ is consistent with local estimates from
bulge mass and velocity dispersion functions
\citep[e.g.,][]{Marconi04,Shankar04}, with room for observationally
estimated Compton-thick fractions of $\sim30-40\%$ which we have included 
(but not, e.g.\ $\gtrsim60\%$). The predicted X-ray
background spectrum agrees well with that observed, in a similar
manner to that in \citet{Ueda03}.  The UV luminosity density and
ionization rate are consistent with all existing constraints, and
account for observed ionization rates from \citet{Scott00} at
$z\lesssim1$ and a significant ($\sim50\%$) fraction at $z\sim2-3$,
but subsequently dropping more rapidly than observed rates. Because of
the flattening of the bright and faint-end QLF slopes with redshift,
the luminosity density at high redshift is dominated by bright
quasars, and above $z\sim2.5$ the number density of bright, optical
quasars (e.g.\ $M_{B}<-27$) becomes an accurate proxy for the UV
luminosity density, related by a conversion which we calibrate (see
\S~\ref{sec:UV}).

We can also de-convolve the bolometric QLF with the quasar lifetime as
a function of luminosity and black hole mass; similar to the procedure
in e.g.\ \citet{YT02} but adopting the physically motivated quasar
lifetimes from \citet{H06a,H06b} to determine the ``birthrate''
(formation rate) of black holes as a function of mass and redshift.
The resulting evolution of the black hole mass function illustrates
the feedback-driven scenario above, with rapid buildup to $z\sim3$,
then activity shutting down in the highest mass systems, and the
characteristic mass of black holes ``in formation'' shifting from
$\sim10^{9}\,M_{\sun}$ at $z\sim2$ to $\sim2\times10^{7}\,M_{\sun}$ at
$z\sim0$. The resulting black hole mass function agrees well with that
inferred locally \citep{Marconi04,Shankar04}.

Measurements of the QLF are converging towards a self-consistent
picture of the QLF shape and evolution across $\sim8$ orders of
magnitude in luminosity and $\sim9$ in space density, frequencies from
$\sim24\,\mu{\rm m}$ to $\sim100\,$keV, through mid and near-IR, optical, UV,
soft and hard X-rays. Except at the highest redshifts $z\gtrsim4$, where statistics
are still poor and faint luminosities have not been probed, 
systematic errors will almost certainly overwhelm statistical errors.
Indeed, comparison of samples with overlapping redshift and luminosity
intervals shows that while the agreement is good, the systematic
offsets (albeit small) between the largest samples are much larger
than their typical statistical errors. Some of these effects, such as
cosmic variance, are at least straightforward to understand and
reduce. However, others are more fundamental. It may not be possible
to define a bolometric QLF to any greater accuracy than we have done
(again, excepting the highest redshifts where larger samples would
improve things considerably), at least in the near future.

For example, on both small scales characteristic of accretion disks,
black hole driven winds, and molecular outflows or tori, and large
scales characteristic of potential merger-driving and galaxy-scale
quasar obscuration, quasars are not isotropic. So even a complete
spectrum does not imply the bolometric luminosity, lacking a more
complete model for the spectrum as a function of viewing angle and
other quantities (e.g.\ accretion rate, mass, spin).  Perhaps by
considering the {\em distribution} of luminosities and accounting for
the fact that there is not one spectral shape for a given total
bolometric luminosity, we have statistically accounted for some of
this, but it is by no means clear, and will not necessarily be so if
there are correlations between e.g.\ the column density distribution
and intrinsic Eddington ratio or spectral shape distribution, or if
e.g.\ the dependence of spectral shape on luminosity is really driven
by an accretion rate or viewing angle dependence.  These are important
but difficult and often degenerate questions, which can only be
answered with more detailed observations of quasar spectra at many
different frequencies, coupled with detailed modeling of accretion
processes and quasar fueling.

Finally, in order to enable simple comparison of our results with 
future theoretical models and 
observations at a wide variety of wavelengths and redshifts, 
we provide for public use\footnote{\calculatorurl} a 
simple ``QLF calculator'' script to return the QLF at a given redshift, 
in an arbitrary observed band or frequency. This calculates the observed QLF
using the full modeling of bolometric corrections and 
extinction discussed herein, for any of the QLF fits in 
Tables~\ref{tbl:params} \& \ref{tbl:params.ldde} or for an arbitrary 
bolometric QLF. As we have 
emphasized for the large number of historical 
fits to the QLF, {\em the fits therein should not be extrapolated beyond 
the observed ranged in Figures~\ref{fig:show.z.low} \&\ \ref{fig:show.z.high}}, 
for example to very high redshifts ($z\gtrsim6$), low luminosities at 
high redshifts ($z\gtrsim4.5$), and unconstrained wavelength regimes 
such as the extreme IR, UV, or hard X-rays.

\acknowledgments We thank G\"{u}nther Hasinger for 
helpful discussions, as well as T.~J.~Cox, Brant Robertson, 
and Scott Tremaine.  We are also grateful to an anonymous
referee for comments that improved this paper.
This work was supported in part by NSF grants ACI
96-19019, AST 00-71019, AST 02-06299, and AST 03-07690, and NASA ATP
grants NAG5-12140, NAG5-13292, and NAG5-13381. GTR further acknowledges 
support from a Gordon and Betty Moore Fellowship in data intensive sciences.

\clearpage
\lscapeopen
\begin{deluxetable}{llcccccc}
\rotator
\sizer
\tablecaption{Measurements of the QLF\label{tbl:qlfs}}
\tablewidth{0pt}
\tablehead{
\colhead{Reference} &
\colhead{Survey/Field\tablenotemark{a}} &
\colhead{Rest Wavelength/Band} &
\colhead{$z$ Range\tablenotemark{b}} &
\colhead{Luminosity Range\tablenotemark{b}} &
\colhead{$\reducechi$\tablenotemark{c}} & 
\colhead{$N_{\rm AGN}$} & 
\colhead{Plotting Symbol}
}
\startdata
Optical: & & & & & & & \\
\hline  \\
Cristiani et al.~(2004) & GOODS & $1450$\,\AA\ & $\sim 4-5.2$ & 
$-21>M_{1450}>-23.5$ & 0.58/1 & $1-4$ & crosses \\
Croom et al.~(2004) & 2QZ/6QZ & $B$ & $0.4-2.1$ & 
$-20.5>M_{g}>-28.5$ & 23.1/10 & 20,905 & asterisks \\
Fan et al.~(2001a) & SDSS (Equatorial Stripe) & $1450$\,\AA\ & $3.6-5.0$ & 
$-25.5>M_{1450}>-27.5$ & 6.21/9 & 39 & pentagons \\
Fan et al.~(2001b,\,2003,\,2004) & SDSS (Main \& Southern Survey) & $1450$\,\AA\ & $\sim5.7-6.4$ & 
$-26.5>M_{1450}>-28$ & 2.12/3 & 9 & ... \\
Hunt et al.~(2004) & LBG survey & $1450$\,\AA\ & $\sim 2-4$ & 
$-21>M_{1450}>-27$ & 4.74/6 & 11 & diamonds \\
Kennefick et al.~(1995) & $POSS$ & $B$ & $4.0-4.5$ & 
$-26.5>M_{B}>-28.5$ & 14.8/2 & 10 & triangles \\
Richards et al.~(2005) & 2dF-SDSS & $g$ & $0.3-2.2$ & 
$-21>M_{g}>-27$ & 137/99 & 5,645 & circles \\
Richards et al.~(2006b) & SDSS (DR3) & $i(z=2)\sim2500$\,\AA\ & $0.3-5.0$ & 
$-22.5>M_{i}>-29$ & 247/101 & 15,343 & squares \\
Schmidt et al.~(1995) & $PTGS$ & $B$ & $\sim 3.5-4.5$ & 
$-25.5>M_{B}>-27.5$ & 8.04/4 & 8 & inverted triangles \\
Siana et al.~(2006) & SWIRE (ELIAS-N1/N2) & $1450$\,\AA\ & $\sim 2.8-3.4$ & 
$-23.5>M_{1450}>-26.5$ & 4.74/6 & $\sim$100 & crosses \\
Wolf et al.~(2003) & COMBO-17 & $1450$\,\AA\ & $1.2-4.8$ & 
$-23.5>M_{1450}>-28.5$ & 54.2/27 & 192 & stars \\
\\
Soft X-ray: & & & & & & & \\
\hline  \\
Hasinger et al.~(2005) & $ROSAT$ (RASS+RDS) + CDF-N/S & $0.5-2$\,keV & $0.015-4.8$ & 
$10^{42}<L_{0.5-2}<10^{48}\,{\rm erg\,s^{-1}}$ & 169/51 & 2,566 & circles \\
Miyaji et al.~(2000,\,2001) & $ROSAT$ (RASS+RDS) & $0.5-2$\,keV & $0.015-4.8$ & 
$10^{41}<L_{0.5-2}<10^{47}\,{\rm erg\,s^{-1}}$ & 112/41 & 691 & stars \\
Silverman et al.~(2005b) & CHAMP+$ROSAT$ (RASS) & $0.5-2$\,keV & $0.1-5$ & 
$10^{44.5}<L_{0.5-2}<10^{46}\,{\rm erg\,s^{-1}}$ & 24.1/9 & 217 & squares \\
 \\
Hard X-ray: & & & & & & & \\
\hline  \\
Barger et al.~(2003a,b) & CDF-N & $2-8$\,keV & $\sim5-6.5$ & 
$10^{43}<L_{2-8}<10^{45}\,{\rm erg\,s^{-1}}$ & 1.02/1 & 1 & diamonds \\
Barger et al.~(2005) & CDF-N/S + CLASXS + $ASCA$ & $2-8$\,keV & $\sim0.1-1.2$ & 
$10^{42}<L_{2-8}<10^{46}\,{\rm erg\,s^{-1}}$ & 41.0/30 & 601 & squares \\
\ ... & CDF-N/S + CLASXS & $2-8$\,keV & $\sim1.5-5.0$ & 
$10^{42}<L_{2-8}<10^{46}\,{\rm erg\,s^{-1}}$ & 15.5/9 & $\sim$100 & ... \\
Barger \&\ Cowie (2005) & CDF-N/GOODS-N & $2-8$\,keV & $\sim2-3$ & 
$10^{43}<L_{2-8}<10^{44.5}\,{\rm erg\,s^{-1}}$ & 1.73/1 & 136 & ... \\
La Franca  et al.~(2005) & HELLAS2XMM & $2-10$\,keV & $0.0-4.0$ & 
$10^{42}<L_{2-10}<10^{46.5}\,{\rm erg\,s^{-1}}$ & 14.4/18 & 508 & stars \\
Nandra et al.~(2005) & GWS + HDF-N & $2-10$\,keV & $2.7-3.2$ & 
$10^{43}<L_{2-10}<10^{44.5}\,{\rm erg\,s^{-1}}$ & 0.77/1 & 15 & crosses \\
Sazonov \&\ Revnivtsev (2004) & RXTE & $3-20$\,keV & $0.0-0.1$ & 
$10^{41}<L_{3-20}<10^{46}\,{\rm erg\,s^{-1}}$ & 9.75/10 & 77 & inverted triangles \\
Silverman et al.~(2005a,c) & CHAMP & $0.3-8.0$\,keV & $0.2-4.0$ & 
$10^{42}<L_{0.3-8}<10^{45.5}\,{\rm erg\,s^{-1}}$ & 26.3/15 & 368 & triangles \\
Ueda et al.~(2003) & $HEAO\,1$ + AMSS-n/s + ALSS & $2-10$\,keV & $0.015-3.0$ & 
$10^{41.5}<L_{2-10}<10^{46.5}\,{\rm erg\,s^{-1}}$ & 26.5/35 & 247 & circles \\
& \ \ \ \ \ + $ASCA$ + CDF-N  & & & & & & \\
 \\
Mid-IR: & & & & & & & \\
\hline  \\
Brown et al.~(2006) & {\em Spitzer} Bo\"{o}tes (NDWFS) & $8\,\mu{\rm m}$ & $\sim1-5$ & 
$10^{45}<L_{8\,\mu{\rm m}}<10^{47}\,{\rm erg\,s^{-1}}$ & 3.77/10 & 183 & circles \\
Matute et al.~(2006) & RMS + ELIAS + HDF-N/S & $15\,\mu${\rm m} & $\sim0.1-1.2$ & 
$10^{42}<L_{15\,\mu{\rm m}}<10^{47}\,{\rm erg\,s^{-1}}$ & 23.4/18 & 148 & squares \\
 \\
Emission Lines: & & & & & & & \\
\hline  \\
Hao et al.~(2005) & SDSS (main galaxy sample) & H$\alpha$ & $0-0.33$ & 
$10^{5}<L_{\rm H\alpha}<10^{9}\,L_{\sun}$ & 29.5/21 & $\sim$3000 & pentagons \\
\ ... & \ ... & [O{\small II}] & \ ... & 
$10^{5}<L_{\rm O\,II}<10^{8}\,L_{\sun}$ & ... & \ ... & ... \\
\ ... & \ ... & [O{\small III}] & \ ... & 
$10^{5}<L_{\rm O\,III}<10^{9}\,L_{\sun}$ & ... & \ ... & ... \\
\enddata
\tablenotetext{a}{For a detailed description of each sample, we 
direct the reader to the listed references (and references therein).}
\tablenotetext{b}{Redshift and luminosity ranges listed are for the 
{\em entire} sample in each case, they should not be taken to imply that 
the observations simultaneously span both ranges.}
\tablenotetext{c}{Reduced $\chi^{2}$ of binned QLF with 
respect to our full best-fit.}
\end{deluxetable}
\clearpage

\clearpage
\pagestyle{empty}
\begin{deluxetable}{cccccc}
\tablecaption{Best-Fit QLF at Various Redshifts\label{tbl:fits.at.z}}
\tablewidth{-264.86507pt}
\tablehead{
\colhead{$\langle z \rangle$} &
\colhead{$\log\phistar$\tablenotemark{a}} &
\colhead{$\log\lstar$\tablenotemark{b}} &
\colhead{$\slopefaint$} &
\colhead{$\slopebright$} & 
\colhead{$\reducechi$}  
}
\startdata
0.1 & $-5.45\pm0.28$ & $11.94\pm0.21$ & $0.868\pm0.050$ & $1.97\pm0.17$ & $89/73$ \\
0.5 & $-4.66\pm0.26$ & $12.24\pm0.18$ & $0.600\pm0.136$ & $2.26\pm0.23$ & $124/66$ \\
1.0 & $-4.63\pm0.15$ & $12.59\pm0.11$ & $0.412\pm0.122$ & $2.23\pm0.15$ & $182/69$ \\
1.5 & $-4.75\pm0.19$ & $12.89\pm0.13$ & $0.443\pm0.145$ & $2.29\pm0.20$ & $214/86$ \\
2.0 & $-4.83\pm0.05$ & $13.10\pm0.04$ & $0.320\pm0.046$ & $2.39\pm0.07$ & $66/67$ \\
2.5 & $-4.96\pm0.14$ & $13.13\pm0.09$ & $0.302\pm0.091$ & $2.30\pm0.15$ & $72/53$ \\
3.0 & $-5.23\pm0.12$ & $13.17\pm0.10$ & $0.395\pm0.060$ & $2.10\pm0.12$ & $45/53$ \\
4.0 & $-4.66\pm0.37$ & $12.39\pm0.32$ & -$0.254\pm0.736$ & $1.69\pm0.18$ & $54/32$ \\
5.0 & $-5.38\pm1.19$ & $12.46\pm1.10$ & $0.497\pm0.458$ & $1.57\pm0.41$ & $14/13$ \\
6.0 & $-5.13\pm0.38$ & $11.0$ & $0.0$ & $1.11\pm0.13$ & $5/3$ \\
\enddata
\tablenotetext{a}{${\rm Mpc^{-3}}$}
\tablenotetext{b}{$L_{\sun}\equiv 3.9\times10^{33}\,{\rm erg\,s^{-1}}$}
\end{deluxetable}

\begin{deluxetable}{lcccccccccccc}
\rotator
\sizer
\tablecaption{Best-Fit QLF to all Redshifts\label{tbl:params}}
\tablewidth{0pt}
\tablehead{
\colhead{Model} &
\colhead{$\log\phistar$\tablenotemark{a}} &
\colhead{($\log\lstar)_{0}$\tablenotemark{b}} &
\colhead{$k_{L,\,1}$} &
\colhead{$k_{L,\,2}$} & 
\colhead{$k_{L,\,3}$} &   
\colhead{$(\slopefaint)_{0}$} &
\colhead{$k_{\slopefaint}$} &   
\colhead{$(\slopebright)_{0}$} &
\colhead{$k_{\slopebright,\,1}$} &   
\colhead{$k_{\slopebright,\,2}$} & 
\colhead{$\rho_{\rm BH}(z=0)$\tablenotemark{c}} & 
\colhead{$\reducechi$} 
}
\startdata
PLE & $-4.733\pm0.101$ & $12.965\pm0.074$ & $0.749\pm0.084$ & $-8.03\pm0.35$ & $-4.40\pm1.05$ 
& $0.517\pm0.065$ & 0 & $2.096\pm0.083$ & 0 & 0 
& $5.66$ & $1924/511$ \\
Faint & $-4.630\pm0.075$ & $12.892\pm0.058$ & $0.717\pm0.069$ & $-8.10\pm0.29$ & $-3.90\pm0.85$ 
& $0.272\pm0.073$ & $-0.972\pm0.268$ & $2.048\pm0.063$ & 0 & 0 
& $4.97$ & $1422/510$ \\
Bright & $-4.930\pm0.070$ & $13.131\pm0.051$ & $0.360\pm0.095$ & $-11.63\pm0.48$ & $-10.68\pm1.04$ 
& $0.605\pm0.044$ & 0 & $2.350\pm0.087$ & $1.53\pm0.14$ & $-0.745\pm0.081$ 
& $5.47$ & $1312/509$ \\
Scatter\tablenotemark{d} & $-4.815\pm0.040$ & $13.064\pm0.030$ & $0.356\pm0.061$ & $-11.69\pm0.27$ & $-9.18\pm0.58$ 
& $0.351\pm0.032$ & $-0.826\pm0.092$ & $2.359\pm0.052$ & $1.534\pm0.070$ & $-0.889\pm0.042$ 
& $4.63$ & $445/508$ \\ 
Schechter\tablenotemark{e} & $-3.579\pm0.092$ & $11.482\pm0.162$ & $-1.78\pm0.29$ & $-21.22\pm1.62$ & $-25.93\pm3.08$ 
& $0.013\pm0.011$ & $-3.760\pm0.859$ & $0.354\pm0.020$ & $1.794\pm0.087$ & $-0.784\pm0.063$ 
& $1.16$ & $1254/507$ \\
Full & $-4.825\pm0.060$ & $13.036\pm0.043$ & $0.632\pm0.077$ & $-11.76\pm0.38$ & $-14.25\pm0.80$ 
& $0.417\pm0.055$ & $-0.623\pm0.132$ & $2.174\pm0.055$ & $1.460\pm0.096$ & $-0.793\pm0.057$ 
& $4.81$ & $1007/508$ \\
\enddata
\tablenotetext{a}{${\rm Mpc^{-3}}$}
\tablenotetext{b}{$L_{\sun}\equiv 3.9\times10^{33}\,{\rm erg\,s^{-1}}$}
\tablenotetext{c}{${\rm 10^{5}\,M_{\sun}\,Mpc^{-3}}$}
\tablenotetext{d}{This fit allows for an intrinsic sample-to-sample systematic normalization 
variance of $0.05$\,dex. However, this was applied uniformly to the samples of 
Table~\ref{tbl:qlfs}, and under-weights much of the most well-constrained data in 
comparison to the ``Full'' fit, which has a higher $\chi^{2}/\nu$ but more faithfully 
represents those data sets.}
\tablenotetext{e}{Adopting the modified Schechter function fit of Equation~(\ref{eqn:expshape})
instead of a double power law.}
\tablenotetext{  \   }{ }
\tablenotetext{  \   }{The resulting observed QLF at arbitrary frequency and redshift 
from any of these fits can be calculated using the QLF calculator script 
available for download at \\ 
\calculatorurl}
\end{deluxetable}

\begin{deluxetable}{lccccccccccccc}
\rotator
\sizer
\tablecaption{Best-Fit LDDE QLF\label{tbl:params.ldde}}
\tablewidth{0pt}
\tablehead{
\colhead{Model} &
\colhead{$\log\phistar$\tablenotemark{a}} &
\colhead{$\log\lstar$\tablenotemark{b}} &
\colhead{$\slopefaint$} &
\colhead{$\slopebright$} &
\colhead{$\log L_{c}$\tablenotemark{b}} &
\colhead{$z_{c\,0}$} &
\colhead{$\alpha$} &
\colhead{$p1_{46}$} &
\colhead{$p2_{46}$} &
\colhead{$\beta_{1}$} &
\colhead{$\beta_{2}$} &
\colhead{$\rho_{\rm BH}(z=0)$\tablenotemark{c}} & 
\colhead{$\reducechi$} 
}
\startdata
PDE & $-6.66\pm0.28$ & $46.64\pm0.19$ & $0.858\pm0.075$ & $2.09\pm0.20$ & $46$ 
& $1.852\pm0.068$ & $0$ & $4.13\pm0.29$ & $-2.53\pm0.46$ & $0$ & $0$ 
& $9.78$ & $3255/511$ \\
LDDE & $-6.20\pm0.15$ & $45.99\pm0.10$ & $0.933\pm0.045$ & $2.20\pm0.14$ & $46.72\pm0.05$ 
& $1.852\pm0.025$ & $0.274\pm0.025$ & $5.95\pm0.23$ & $-1.65\pm0.21$ & $0.29\pm0.34$ & $-0.62\pm0.17$ 
& $5.09$ & $1389/507$ \\
\enddata
\tablenotetext{a}{${\rm Mpc^{-3}}$}
\tablenotetext{b}{${\rm erg\,s^{-1}}$}
\tablenotetext{c}{${\rm 10^{5}\,M_{\sun}\,Mpc^{-3}}$. In this formulation, the luminosity density 
is nearly divergent at low $L$, so the integration is truncated at $10^{41}\,{\rm erg\,s^{-1}}$.}
\tablenotetext{  \   }{ }
\tablenotetext{  \   }{The resulting observed QLF at arbitrary frequency and redshift 
from any of these fits can be calculated using the QLF calculator script 
available for download at \\
\calculatorurl}.
\end{deluxetable}

\begin{deluxetable}{lccccccccccc}
\rotator
\sizer
\tablecaption{Best-Fit Quasar Formation Rate\label{tbl:params.deconvolved}}
\tablewidth{0pt}
\tablehead{
\colhead{Model} &
\colhead{$(\log\dot{\phi}_{\ast})_{0}$\tablenotemark{a}} &
\colhead{$k_{\dot{\phi}}$} &
\colhead{($\log{M_{\ast}})_{0}$\tablenotemark{b}} &
\colhead{$k_{M,\,1}$} &
\colhead{$k_{M,\,2}$} & 
\colhead{$\eta_{1}$\tablenotemark{c}} &
\colhead{$(\eta_{2})_{0}$} &
\colhead{$k_{\eta_{2},\,1}$} &   
\colhead{$k_{\eta_{2},\,2}$} & 
\colhead{$\rho_{\rm BH}(z=0)$\tablenotemark{d}} & 
\colhead{$\reducechi$} 
}
\startdata
Fixed $\slopebright$ & $-3.808\pm0.037$ & $-4.46\pm0.31$ & $8.894\pm0.037$ 
& $1.54\pm0.12$ & $-4.46\pm0.31$ & $0.20$ 
& $2.44\pm0.11$ & $0$ & $0$ 
& $6.19$ & $1338/511$ \\
Variable $\slopebright$ & $-3.830\pm0.031$ & $-4.02\pm0.36$ & $8.959\pm0.032$ 
& $1.18\pm0.13$ & $-6.68\pm0.44$ & $0.20$ 
& $2.86\pm0.16$ & $1.80\pm0.18$ & $-1.13\pm0.09$ 
& $6.28$ & $1206/509$ \\
Schechter\tablenotemark{e} & $-3.571\pm0.080$ & $-3.81\pm0.46$ & $8.710\pm0.130$ 
& $1.33\pm0.21$ & $-7.64\pm0.99$ & $0.20$ 
& $0.79\pm0.10$ & $1.33\pm0.19$ & $-1.17\pm0.12$ 
& $4.27$ & $1434/509$ \\
\enddata
\tablenotetext{a}{${\rm Mpc^{-3}\,Gyr^{-1}}$}
\tablenotetext{b}{$M_{\sun}$}
\tablenotetext{c}{In this model, the QLF faint-end slope is dominated 
by the shape of the quasar lightcurve, giving only weak constraints on $\eta_{1}$. We 
adopt the maximum acceptable $\eta_{1}\approx0.20$ ($1\sigma$) as an upper limit, 
and show the fit results for this case.}
\tablenotetext{d}{${\rm 10^{5}\,M_{\sun}\,Mpc^{-3}}$}
\tablenotetext{e}{Adopting the modified Schechter function fit of Equation~\ref{eqn:expshape}
instead of a double power law.}
\end{deluxetable}

\pagestyle{headings}

\clearpage
\lscapeclose

\end{document}